\documentclass[fleqn,usenatbib,onecolumn]{mnras}

\usepackage{newtxtext,newtxmath}
\usepackage[T1]{fontenc}

\DeclareRobustCommand{\VAN}[3]{#2}
\let\VANthebibliography\thebibliography
\def\thebibliography{\DeclareRobustCommand{\VAN}[3]{##3}\VANthebibliography}

\usepackage{graphicx}	% Including figure files
\usepackage{amsmath}	% Advanced maths commands
\usepackage{url}            % simple URL typesetting
\usepackage{nicefrac}       % compact symbols for 1/2, etc.
\usepackage{microtype}      % microtypography
\usepackage{lipsum}		% Can be removed after putting your text content
\usepackage{graphicx}
\usepackage{natbib}
\usepackage{bbold}
\usepackage{subfig}
\usepackage{mwe}
\usepackage{verbatim}
\usepackage{titlesec}
\usepackage{float}

\newcommand{\D}{\mathrm{d}}
\newcommand{\E}{\mathrm{e}}
\newcommand{\I}{\mathrm{i}}
\newcommand{\tens}[1]{\boldsymbol{#1}}

\title[Kinetic Field Theory Applied to Planetesimal Formation I]{Kinetic Field Theory Applied to Planetesimal Formation I: Freely Streaming Dust Particles}

\author[J. Shi et al.]{
Jiahan Shi,$^{2,3}$\thanks{E-mail: shi@mpia.de}, 
Matthias Bartelmann$^1$,
Hubert Klahr$^2$,
Cornelis P.\ Dullemond$^3$
\\
$^{1}$Institut für Theoretische Physik, Universität Heidelberg, Germany\\
$^{2}$Max Planck Institut für Astronomie, Heidelberg, Germany\\
$^{3}$Zentrum für Astronomie, Universität Heidelberg, Germany
}

\date{Accepted XXX. Received YYY; in original form ZZZ}

\pubyear{2024}

\begin{document}
\label{firstpage}
\pagerange{\pageref{firstpage}--\pageref{lastpage}}
\maketitle

% Abstract of the paper
\begin{abstract}
Planet formation in the solar system was started when the first planetesimals were formed from the gravitational collapse of pebble clouds.
Numerical simulations of this process, especially in the framework of streaming instability, produce various power laws for the initial mass function for planetesimals. While recent advances have shed light on turbulence and its role in particle clustering, a comprehensive theoretical framework linking turbulence characteristics to particle cluster properties and planetesimal mass function remains incomplete.
Recently, a kinetic field theory for ensembles of point-like classical particles in or out of equilibrium 
has been applied to cosmic structure formation. This theory encodes the dynamics of a classical particle ensemble by a generating functional specified by the initial probability distribution of particles in phase space and their equations of motion. Here, we apply kinetic field theory to planetesimal formation. A model for the initial probability distribution of dust particles in phase space is obtained from a quasi-initial state for a three-dimensional streaming-instability simulation that is a particle distribution with velocities for gas and particles from the Nakagawa relations. The equations of motion are chosen for the simplest case of freely streaming particles. We calculate the non-linearly evolved density power spectrum of dust particles and find that it develops a universal $k^{-3}$ tail at small scales, suggesting scale-invariant structure formation below a characteristic and time-dependent length scale. Thus, the KFT analysis indicates that the initial state for streaming instability simulations does not impose a constraint on structure evolution during planetesimal formation.
\end{abstract}

% Select between one and six entries from the list of approved keywords.
% Don't make up new ones.

\section{Introduction}

Planet formation starts with the coagulation of cosmic dust particles in a protoplanetary disk. Dust aggregates grow to ever increasing sizes, but the growth tends to stop once particles reach either the ``fragmentation barrier'' or the ``drift barrier''. The fragmentation barrier is the grain size range at which collisions become destructive \citep{refId0, blum2008, Güttler2010}, while the radial drift barrier is the grain size range at which particles radially drift faster than they can grow \citep{Weidenschilling1977, refId0}. The typical dust aggregate size at which these barriers prevent further growth is of the order of decimeters at 1 au from the star, or millimeters at tens to hundreds au from the star. Although the drift barrier can be overcome through dust traps \citep{Whipple1972, Kretke_2007}, which are now observed to be abundant in protoplanetary disks \citep{Brogan_2015, andrews_2018}, it remains unclear if the fragility of the van der Waals sticking force between macroscopic ($\gtrsim 1$ mm size) dust aggregates allows growth beyond a few centimeters. It is therefore highly questionable if pure dust coagulation can explain the formation of planets.

It has, however, long been known that dust particles can cluster in passive turbulent gas flows \citep[e.g.][]{1997JFM...335...75S, PhysRevE.60.1674, pan_2011}. Since protoplanetary disks are thought to be more or less turbulent, this process is likely to occur in these environments. In addition to passive particle concentration, in which the particles concentrate in a given background turbulence of the gas, it turns out that at high enough initial dust-to-gas density ratios, the particles can also feed back onto the gas, and {\em induce} turbulence, which then leads to a self-sustained cycle of turbulence and particle concentration in a single model \citep{youdin_2005, youdin_2007, johansen_2007}.  This process, called the ``streaming instability'' (SI), is currently the most promising candidate for this particle concentration mechanism. If particle concentrations reach densities larger than the Roche density, they can gravitationally contract, collapse, and form planetesimals \citep[e.g.][]{ nature2007, Cuzzi_2008,  CUZZI2010518}.  A promising scenario for planet formation is therefore that dust grows initially through coagulation, until the particles are large enough to start concentrating in the turbulent gas flow, but still small enough to avoid fragmentation or excessive radial drift. If the concentration becomes strong enough, gravitational collapse of the particle cloud forms a planetesimal on a dynamic time scale, and these planetesimals subsequently gravitationally agglomerate, on a much larger time scale, to form planets. In this way, the barriers can be elegantly ``jumped over''. This scenario appears to also predict the right size distributions of solar system minor bodies \citep[e.g.][]{MORBIDELLI2009558}.

However, there are still numerous problems. For instance, the SI works best for particles with Stokes numbers close to unity, which is also the particle size regime where fragmentation is most destructive. Particles small enough to fall below the fragmentation barrier have Stokes numbers substantially less than unity, which is where the streaming instability is weakened or not operative.  \citet{yang2017} found that the SI can still produce planetesimals starting from particles with Stokes numbers down to $10^{-3}$, as long as the dust density is high enough. But these simulations require very high resolution and long computation time. In general, the smaller the Stokes number, the more spatial resolution the simulations require.

In the last few years, simulations of the streaming instability were done with a particle size distribution instead of a single size for all dust particles \citep[e.g.][]{Krapp_2019, Paardekooper2020, zhu2020, mcnally2021}. It was found that the behavior of the streaming instability under these more realistic conditions is fundamentally different than for the monodisperse case. These simulations require even more computational power. The cost of 3D simulations scales with $\Delta x^{-4}$ (one factor for each spatial dimension and one for time), where $\Delta x$ is the grid scale. For each factor of 2 improvement in resolution, the cost increases by a factor of 16. As a result, numerical simulations of particle clustering in turbulent gas flows are typically limited by what one can computationally afford, and it is hard to know if the affordable resolution is sufficient.

To overcome these problems, various techniques have been developed over the years to treat the particle concentration problem (semi-)analytically, including scaling considerations \citep[e.g.][]{hogen, hopkins, hartlep2017, Hartlep_2020}, and particle simulations in semi-analytical turbulence \citep[e.g.][]{hubbard2013, booth2017}. The advantage of these methods is that they are not (or at least: less) limited by computational resources, and they can therefore handle much larger dynamic ranges.
%\cpdnote{Check for later literature, e.g. Hartlep & Cuzzi 2017 or maybe more recent; perhaps perform an inverse-ADS-search starting from the Hogan, Hopkins, Hartlep papers} 

The goal of this paper is to explore the feasibility of applying an entirely new class of analytic modeling methods to the problem of particle clustering: Kinetic Field Theory (KFT), with initial conditions taken from numerical simulations.

Our KFT approach can be considered to be an extension of the work by e.g. \cite{hogen} in the sense that they investigate the creation of overdense clumps for the case of Kolmogorov turbulence. The ultimate goal of such a project is to derive a prediction for the density probability distribution function (PDF) based on a KFT analysis of the "large" scale structure of realistic disk turbulence as is shown in simulations of streaming instabilities, as well as magnetic and thermodynamic processes driving turbulence. 
Disk turbulence is not necessarily isotropic, it is heavily influenced by rotation, shear, stratification and a wide range of non-isotropic driving modes. In the present paper we will not aim at deriving a PDF, because we neglect important physical effects, as for instance friction and do not consider a developed turbulent state. But as a ground-laying project we show that A: KFT analysis can be applied to turbulence and that B: the initial condition of SI simulations does not induce a preferred scale for structure formation.

Kinetic Field Theory is a microscopic, non-equilibrium, statistical field theory for initially correlated ensembles of classical microscopic particles as developed in \cite{das2012}, \cite{Bartelmann_2016}, \cite{Bartelmann_2017} and \cite{bartelmann2019cosmic}. It is based upon the Martin-Siggia-Rose approach to classical statistical systems \cite{martin1973} and has been adapted in previous papers to cosmological initial conditions and to the expanding background space-time. KFT describes classical particle ensembles by a generating functional which is completely specified by the statistical properties of the initial state and the Green’s function or propagator of the equations of motion. Unlike other approaches to kinetic theory, KFT is not based on a phase-space density function subject to the Liouville or Boltzmann equations. It is a kinetic theory in the sense that it deals with the joint evolution of a particle ensemble while intentionally integrating over microscopic information. KFT defines the initial state of a particle ensemble by covering phase-space with a probability distribution. It samples this distribution with discrete particles and follows their phase-space trajectories in time. Each trajectory thus carries an initial occupation probability through phase space. The set of all particle trajectories is the field that KFT operates on.

In this first paper we do not aim to solve the problem of particle clustering in turbulent flows yet. Instead we apply it to a simpler problem of particle motion and density fluctuations in a background gas, in which the dust to gas ratio sets the local velocities of gas and particles according to \cite{Nakagawa1986}, with the aim to find out how KFT can be applied to dust clustering in proto-planetary disks.

This paper is structured as follows. In Sec.~\ref{sec:kft}, we review the essentials of KFT. In Sec.~\ref{sec:ipdf}, we show how the initial probability distribution function (iPDF) in phase space can be modelled based on a three-dimensional, streaming-instability simulation of planetesimal formation by Gaussianizing the density and momentum fields. The momentum covariance matrix is estimated from the simulation and assumed to be homogeneous and isotropic. In Sec.~\ref{sec:non_ps}, we specify the particle trajectories. By applying two density operators to the generating functional, we derive the evolved non-linear density power spectrum for freely streaming particles in KFT and find that it necessarily develops a universal $k^{-3}$ tail at small scales. In Sec.~\ref{sec:diss}, we summarise and discuss our results.

We use the convention
\begin{equation}
  \mathcal{F}[f] =: \tilde{f}(\vec{k}) =
  \int_q f(\vec{q})e^{-\I\vec{k}\cdot\vec{q}}\;,\quad
  \mathcal{F}^{-1}[\tilde{f}] =: f(\vec{q}) =
  \int_k\tilde{f}(\vec{k})e^{\I\vec{k}\cdot\vec{q}}
\end{equation}
for the Fourier transform $\mathcal{F}$ and its inverse $\mathcal{F}^{-1}$, abbreviating
\begin{equation}
  \int_q=\int\D^3\Vec{q}\;,\quad
  \int_k=\int\frac{\D^3\Vec{k}}{(2\pi)^3}\;.
\end{equation}

\section{Kinetic Field Theory}
\label{sec:kft}

We now briefly review those parts of KFT which will be essential for our paper. For a more intuitive understanding of the concepts of this theory, we refer the reader to Appendix \ref{app:kft} for a comparison with equilibrium statistical physics.

Kinetic field theory is a statistical field theory for classical particle ensembles in or out of equilibrium. Its central mathematical object is a generating functional $Z$. In close analogy to the partition sum of equilibrium thermodynamics, this generating functional integrates the probability distribution $P(\phi)$ for the system state $\phi$ over the state space,
\begin{equation}\label{eq2.1}
  Z = \int\mathcal{D}\phi P(\phi)\;.
\end{equation}
For classical (canonical) ensembles of $N$ point particles, the state space is the phase space $\Lambda$ of the $N$ particles whose trajectories $(\Vec{q}_j,\Vec{p}_j)=:\Vec{x}_j$ with $1 \leq j \leq N$ are tuples of positions $\Vec{q}_j$ and momenta $\Vec{p}_j$. For a compact notation, we bundle the phase-space trajectories $\Vec{x}_j$ into a tensorial object
\begin{equation}
  \tens x=:\Vec{x}_{j}\otimes \Vec{e}_{j}
\end{equation}
where summation over $j$ is implied, and the vector $\Vec e_j\in \mathbb{R}^N$ has components $(\Vec{e}_j)_i=\delta_{ij}$, $1\leq i \leq N$. For such tensors, we define the scalar product
\begin{equation}\label{eq2.3}
  \left<\tens x,\tens y\right> =
  \left(\Vec{x}_i\cdot \Vec{y}_j\right)\left(\Vec{e}_i\cdot \Vec{e}_j\right) =
  \Vec{x}_j\cdot \Vec{y}_j\;.
\end{equation}

Point particles are described by Dirac delta distributions instead of smooth fields. The path integral in (\ref{eq2.1}) then turns into
\begin{equation}\label{eq2.4}
  Z = \int\mathcal{D}\tens x\,P(\tens x)\;.
\end{equation}

We introduce a generator field $\tens J$ conjugate to the phase space trajectories $\tens x(t)$ into the generating functional $Z$,
\begin{equation}\label{eq2.5}
  Z\to Z[\tens J] = \int\mathcal{D}\tens x\, P(\tens x)\exp\left\{
    \I\int^{\infty}_{0}\D t'\left<\tens J(t'),\tens x(t')\right>
  \right\}
\end{equation}
such that the functional derivative of $Z[\tens J]$ with respect to the component $\Vec{J}_j(t)$ of the generator field returns the average position $\left<\Vec{x}_j(t)\right>$ of particle $j$ at time $t$,
\begin{equation}\label{eq2.6}
  \left<\Vec{x}_j(t)\right> =
  -\I\frac{\delta}{\delta \Vec{J}_j(t)}Z[\tens J]\Bigr\vert_{\tens J=0}\;.
\end{equation}
Like $\tens x$, the generator field has components conjugate to positions $\tens q$ and momenta $\tens p$,
\begin{equation}\label{eq2.7}
  \tens J=\Vec{J}_j\otimes \Vec{e}_j=
  \begin{pmatrix}\Vec{J}_{q_j}\\\Vec{J}_{p_j}\end{pmatrix}
  \otimes\Vec{e}_j\;.
\end{equation}

We further split the probability $P(\tens x)$ for the state $\tens x$ to be occupied into a probability $P(\tens x^{(i)})$ for the particle ensemble to occupy an initial state $\tens x^{(i)}$ at time $t = 0$, times the conditional probability $P(\tens x|\tens x^{(i)})$ for the ensemble to move from there to the time-evolved state $\tens x$,
\begin{equation}\label{eq2.8}
  P(\tens x) = \int\D\tens x^{(i)}\,
  P\left(\tens x|\tens x^{(i)}\right)P\left(\tens x^{(i)}\right)\;.
\end{equation}
For particles on classical trajectories, the transition probability $P(\tens x|\tens x^{(i)})$ must be a functional delta distribution of the classical particle trajectories,
\begin{equation}\label{eq2.9}
  P\left(\tens x|\tens x^{(i)}\right) =
  \delta_\mathrm{D}\left[\tens x-\Phi_{cl}\left(\tens x^{(i)}\right)\right]\;,
\end{equation}
where $\Phi_{cl}(\tens x^{(i)})$ denotes the classical Hamiltonian flow on phase space, beginning at the initial phase-space points $\tens x^{(i)}$ of the particle ensemble. Writing the equation of motion in the form $E(\tens x)=0$, the Hamiltonian flow consists of solutions of this equation for all initial points within a certain domain of phase space.

Denoting the trajectories for the particle ensemble by $\bar{\tens x}$,
\begin{equation}\label{eq2.18}
  \delta_\mathrm{D}\left[\tens x-\Phi_{cl}\left(\tens x^{(i)}\right)\right] =
  \delta_\mathrm{D}\left[\tens x(t)-\bar{\tens x}(t)\right]\;.
\end{equation}
Inserting this expression into (\ref{eq2.9}), the result into (\ref{eq2.8}) and the probability distribution $P(\tens x^{(i)})$ into the generating functional $Z[\tens J] $ from (\ref{eq2.5}) gives
\begin{equation}\label{eq2.19}
  Z[\tens J]=\int\D\Lambda
  \exp\left\{
    \I\int^{\infty}_{0}dt'\left<\tens J(t'),\bar{\tens x}(t')\right>
  \right\}
\end{equation}
where we have introduced the initial phase-space measure
\begin{equation}\label{eq2.20}
  \D\Lambda := \D\tens x^{(i)}P\left(\tens x^{(i)}\right)\;.
\end{equation}

\section{Initial probability distribution from the linear phase of a 3D Streaming-Instability Simulation}\label{sec:ipdf}

In this Section, we analyze the position and velocity data of a three-dimensional streaming-instability simulation for a non-stratified disk without vertical gravity, and derive an initial probability distribution (iPDF) from it according to general criteria. Then, we extract the momentum covariance matrix from the simulation and bring it into a form appropriate for a homogeneous and isotropic system.

\subsection{Specification of the simulation}\label{sim}

The three-dimensional streaming-instability (SI) simulation by the Pencil code, see \cite{BRANDENBURG2002471} and \cite{johansen_2007}, is performed in the shearing-sheet approximation. The simulation volume is covered with a Cartesian coordinate grid co-rotating with the Keplerian orbital frequency $\Omega$, at an arbitrary distance $R_0$ from the central star. The simulation length and time scales are set in proto-planetary disk (PPD) units defined by the scale height $H=H(R_0)$ of the gas disk, and the Keplerian orbital frequency $\Omega$. In the following, all derived quantities, e.g.\ distance and friction coefficients, are thus given in disk units as well. 

A snapshot of the SI simulation has been kindly provided to us by Andreas Schreiber, see Figure 6.4 and Figure C.10 (a) in Chapter 6 of \cite{andreas}. This snapshot, deliberately taken at $t_\mathrm{snap}\ne 0$, is still in a very early linear state, but already contains some structure in its SI typical velocity distribution. We aim at studying what KFT predicts for the further evolution of the power spectrum of these structures. The main simulation parameters are:

 \begin{enumerate}
  \item The friction coefficient between dust particles and gas, $\tau_\mathrm{s}$, as represented by the Stokes number $\mathrm{St}=\tau_\mathrm{s}\Omega=0.01$;
  \item The domain size $L_x=L_y=L_z=L=0.1$ of the simulation in units of the disk scale height $H=c_\mathrm{s}\Omega^{-1}$, where the isothermal sound speed $c_\mathrm{s}$ is taken to be constant. Note that $L$ is chosen large enough for assuming that the structures studied later are statistically homogeneous and isotropic in the simulation domain; 
  
  \item The initial dust-to-gas density ratio, $\varepsilon_0 = \rho_\mathrm{d}/\rho_\mathrm{g} = 1.0$, where the initial mean gas density $\bar{\rho}_g$ is set to $1.0$ for simplicity. This sets the total dust mass to $M_\mathrm{d} = \bar{\rho}_\mathrm{g} L^3 = 10^{-3}$;
  \item The particle number is $N \approx 2.1\cdot10^7$, corresponding to an average of 10 particles per cell for a grid resolution of $128^3$. Particles are initially placed randomly throughout the simulation box, which serves as a seed for streaming instabilities as in \citet{johansen_2007}. Note that in this paper, we are solely interested in the dust particles' behavior, to increase the resolution and retain more information, the particles are in fact analyzed on a $256^3$ grid;
  \item The radial pressure gradient is $\eta = 0.05$, same as in \citet{Klahr_2021}.
  \item The time of the snapshot is $t_\mathrm{snap}=4.25$, which means that this snapshot is taken after the simulation box has orbited the central star $\frac{4.25}{2\pi}$ times. We have thus simulated $4.25$ sound crossing times, in which the flow developed a three-dimensional structure, whereas the initial velocity had no vertical velocity perturbation as follows from the \citet{Nakagawa1986} equations.
  \item For the combination of small $\mathrm{St} = 0.01$ and large box $L = 0.1 H$ we can assure that the flow is at best in the earliest linear phase for SI, as the growth time for the given parameters and the resolution is longer than $10^5$ orbits.
\end{enumerate}

The simulation data is thus in a quasi-equilibrium determined by the initial white noise and the \cite{Nakagawa1986} equations. We therefore capture the most important features of a streaming instability, without having to deal with the non-linear evolution of the system.

\begin{figure}
  \centering
   \includegraphics[width=0.5\hsize]{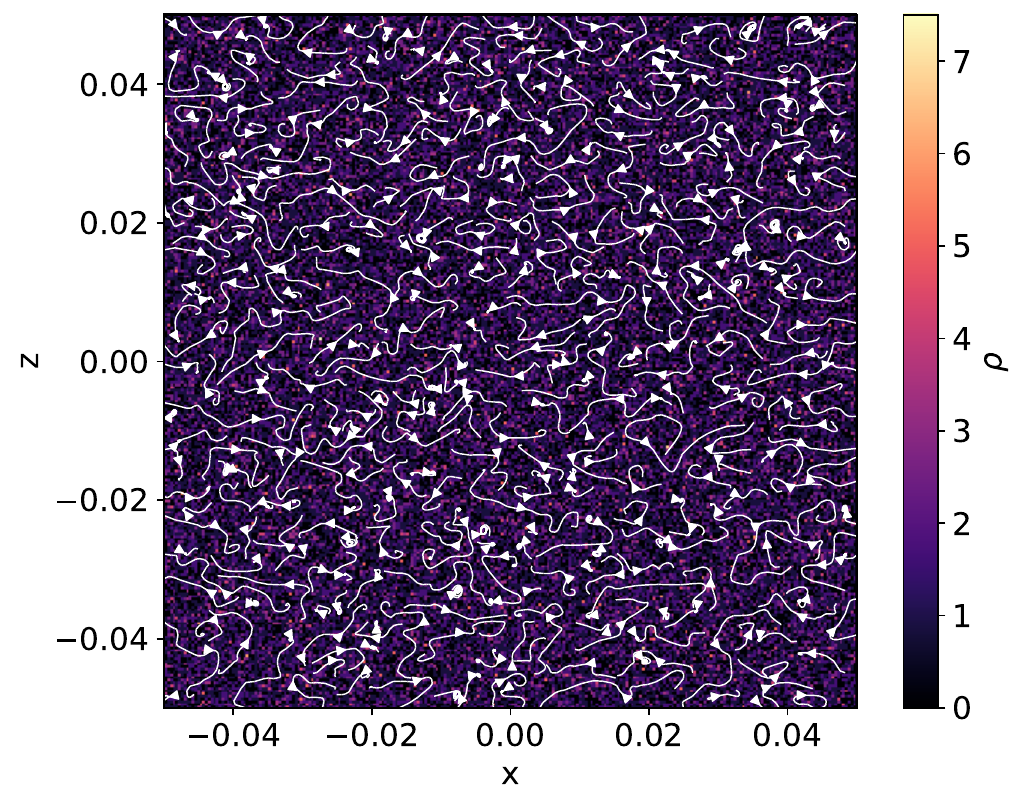}
  \caption{This plot shows the number density $\rho$ and the streamline of dust particles in a $x-z$ slice of the simulation snapshot discussed in this paper. This number density can be well modeled as a statistically isotropic and homogeneous random field.}
\label{fig:density_ill}
\end{figure}

\subsection{Initial probability distribution}
\label{Ini_Pro_Dis}

The initial probability distribution function (iPDF) in phase space is a joint probability density function for the three components of the position vector and the three components of the momentum vector. From the simulation data, it is straightforward to estimate the probability density function for each of these components, but it is quite unfeasible to accurately determine their complete iPDF. We rather choose a different approach: we determine their variances and transform the variables such that we can model their joint distribution by a multivariate Gaussian. The joint probability density function for a $k$-dimensional multivariate Gaussian distribution with mean $\vec\mu$ and covariance $\Sigma$ is
\begin{equation}\label{gaus_joint}
  P(\vec x) = \frac{1}{\sqrt{(2\pi)^k\det\Sigma}}\exp\left[
    -\frac{1}{2}\left(
      \vec x-\vec\mu
    \right)^\top\Sigma^{-1}\left(
      \vec x-\vec\mu
    \right)
  \right]\;,
\end{equation}
where $\vec x$ and $\vec\mu$ are real, $k$-dimensional column vectors.

Since the probability for finding a particle in the simulation is proportional to its density, we only need for variants, viz.\ one for the density and three for the momentum components. If the density and momentum components had themselves Gaussian distributions, we could immediately write down their joint probability distribution as above.

\subsubsection{Density and momentum distributions}
\label{den_mon_dis}

\begin{figure}
  \centering
  \includegraphics[width=0.32\hsize]{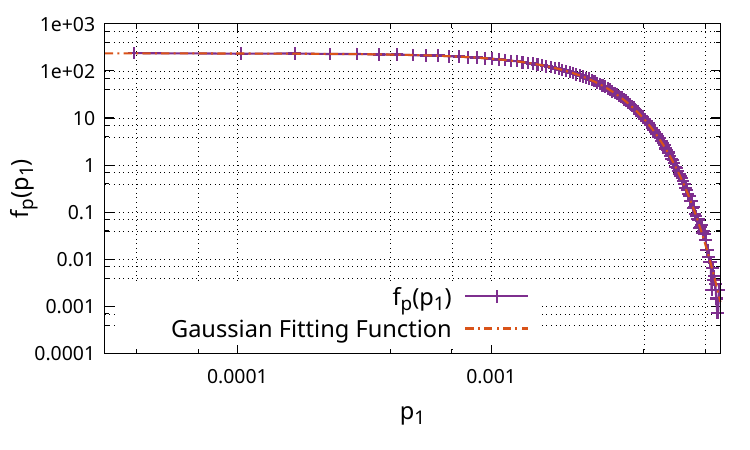}\hfill
  \includegraphics[width=0.32\hsize]{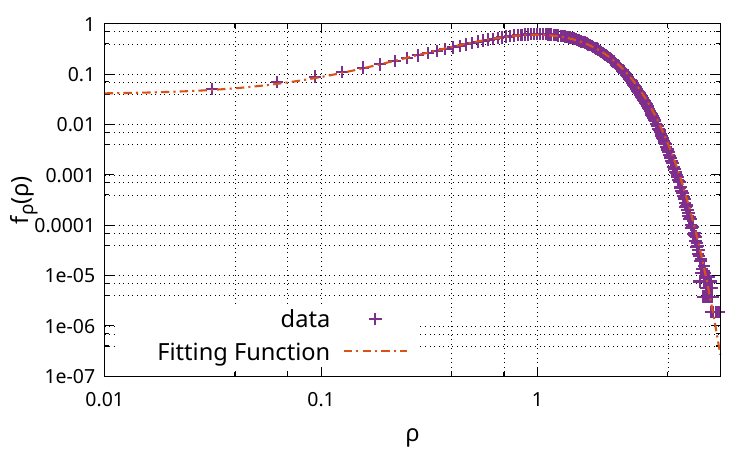}\hfill
  \includegraphics[width=0.32\hsize]{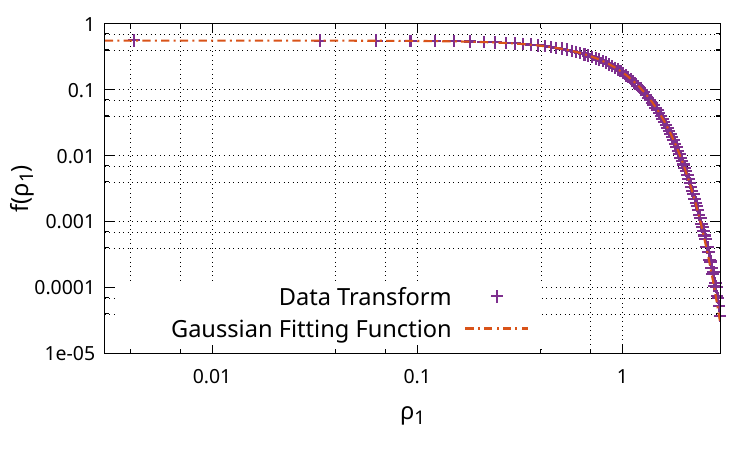}
\caption{\emph{Left}: Momentum distribution in the $x$ direction with double-logarithmic scaling. The purple points represent actual data, the red line shows the Gaussian fit. The mean value $p_{1c}$ has been set to zero. \emph{Centre}: Density distribution function in double-logarithmic scaling. The data points are represented by the purple dots, and the orange curve shows the fit function (\ref{den_dis}). \emph{Right}: Probability distribution of the centred and normalized density variable $\rho_1$ in double-logarithmic scaling. The purple dots show the data points transformed as in (\ref{relation_y_x}). The orange curve shows a Gaussian fit to these points.}
\label{fig:den_dis}
\end{figure}

As an example, the distribution of one of the three momentum components is shown in the left panel of Fig.~\ref{fig:den_dis}. 

All three distributions can individually be fit by Gaussians with amplitudes $A_i$, mean values $p_{ic}$, and standard deviations $\sigma_i$, with $i = x, y, z$. The parameter values for all momentum components are listed in Tab.~\ref{tab_mom}. The non-zero mean momenta in the $x$ and $y$ directions reflect the radial and tangential drifts. Since in KFT, we are exclusively interested in the statistical behavior of the system, we neglect any bulk motion in the following by setting them to zero.

\begin{table}
  \caption{Best-fitting parameters of the momentum distributions in the three spatial directions. The mean values for the momenta in the $x$ and $y$ directions do not vanish due to drift motion. We set them to zero later.}
  \centering
  \begin{tabular}{|c|c|c|c|}
    \hline
      & $A_i$ & $p_{ic}$ & $\sigma_i$ \\
    \hline
    1 & $238.44$ & $-2.36\cdot10^{-4}$ & $1.67\cdot10^{-3}$ \\
    \hline
    2 & $355.40$ & $-2.49\cdot10^{-2}$ & $1.12\cdot10^{-3}$ \\
    \hline
    3 & $413.54$ & $0.0$ & $9.61\cdot10^{-4}$ \\
    \hline
  \end{tabular}
\label{tab_mom}
\end{table}

The density distribution, however, is not a Gaussian; cf.\ Fig.~\ref{fig:den_dis}. 

Rather, it can be approximated by
\begin{equation}
\label{den_dis}
  f_{\rho}(\rho) = \frac{3}{2}\,
  \frac{\left(b+\rho^{3/2}\right)}{b\Gamma(2/3)+\Gamma(5/3)}
  \exp\left(-\rho^{3/2}\right)
\end{equation}
with $b=0.025$, which is normalized to unity in $[0,\infty]$. Here $\Gamma$ is the Gamma function.

\subsubsection{Gaussianization of the density distribution}
\label{mod_den}

Since all momentum components have Gaussian distributions, while the density has not, we transform to a new density variable $\bar\rho$ which is also distributed in a Gaussian way. Let $\bar\rho$ be a strictly monotonic, real-valued function of $\rho$, $\bar\rho = h(\rho)$, then the probability densities $f_\rho$ of $\rho$ and $f_{\bar\rho}$ of $\bar\rho$ are related by
\begin{equation}
  f_{\rho}(\rho) = f_{\bar{\rho}}(h(\rho))h'(\rho)\;,
\end{equation}
where the prime denotes the derivative of $h$ with respect to $\rho$, while the cumulative probability distributions (CDF) must be related by
\begin{equation}
  F_{\rho}(\rho) = F_{\bar{\rho}}(\bar{\rho})\;.
\end{equation}

Since $f_{\bar\rho}$ is supposed to be a Gaussian with mean value $\mu$ and standard deviation $\sigma$, its CDF is the error function
\begin{equation}\label{new_cdf}
  F_{\bar{\rho}}(\bar{\rho}) = \frac{1}{2}\left[
    1+\mathrm{erf}\left(\frac{\bar{\rho}-\mu}{\sigma}\right)
  \right]\;.
\end{equation}
The centred and normalized density variable $\rho_1=(\bar{\rho}-\mu)/\sigma$ is related to the density $\rho$ by
\begin{equation}\label{relation_y_x}
  \rho_1 = \mathrm{erf}^{-1}\left[2F_{\rho}(\rho)-1\right]\;,
\end{equation}
with the CDF of $\rho$ given by
\begin{equation}\label{cdf_rho}
  F_{\rho}(\rho) = \int_0^{\rho} f_{\rho}(\rho)\D\rho =
  \frac{b\gamma\left(2/3,\rho^{3/2}\right)+\gamma\left(5/3,\rho^{3/2}\right)}
       {b\Gamma(2/3)+\Gamma(5/3)}\;.
\end{equation}
Here, $\mathrm{erf}^{-1}$ is the inverse error function, and $\gamma(a,x)$ the incomplete Gamma function.

Both $F_{\rho}(\rho)$ and the inverse error function are monotonically increasing. To illustrate that the new density variable $\rho_1$ does indeed follow a Gaussian distribution, its probability distribution function and a Gaussian fit to it are shown in the right panel of Fig.~\ref{fig:den_dis}.

In the range of densities relevant for us, the function $\rho_1(\rho)$ from (\ref{relation_y_x}) admits the fit
\begin{equation}\label{inverse_func}
  \rho_1 = \frac{\bar\rho-\mu}{\sigma} =
  \sqrt{\frac{\rho}{a_1}}-a_2
\end{equation}
with $a_1 = 0.1823\pm0.0005$ and $a_2 = 2.569\pm0.006$. We can thus choose $\sigma = \sqrt{a_1}$ and $\mu = \sqrt{a_1}a_2$ in (\ref{new_cdf}) to relate the density $\rho$ to the density variable $\bar\rho$ simply by
\begin{equation}\label{inverse_func_rho'}
  \rho = \bar\rho^2\;.
\end{equation}
Thus, the density variable $\bar\rho$, which by construction follows a Gaussian distribution with mean $\mu$ and standard deviation $\sigma$, is well approximated by the square root of the density $\rho$.

\subsubsection{Final expression}

Having assured that the momentum components and the density variable $\bar\rho$ are Gaussian variates to sufficient approximation, we adopt the same method as in Appendix $A$ of \cite{Bartelmann_2016} to find the initial probability distribution function in phase space for dust particles, the final result thus is given by
\begin{equation}\label{final_prob_dist}
  P(\tens q,\tens p) = \frac{N^{-N}\mathcal{C}(-\I\partial_{\tens p})}
  {\sqrt{(2\pi)^{3N}\det\Bar{C}_{pp}}}\exp\left(
    -\frac{1}{2}\tens p^\top\Bar{C}_{pp}^{-1}\tens p
  \right)
\end{equation}
where $\mathcal{C}(-\I\partial_{\tens p})$ given by (\ref{eq:ctp}) is an operator determined by the density-density and density-momentum correlations, and $\Bar{C}_{pp}$ given by (\ref{cov_matr}) is the momentum covariance matrix. The detailed calculation is carried out in Appendix \ref{appa}.

\subsection{Covariance Matrix}\label{Cov_max}

For specifying the joint probability distribution (\ref{final_prob_dist}), we still need the covariance matrix $\bar C_{pp}$, which we model to represent the simulation data. We proceed in four steps:
\begin{enumerate}
  \item \label{Cov_steps1} Place particles on grids, using the cloud-in-cells (CIC) algorithm;
  \item \label{Cov_steps2} Obtain density and momentum fields in Fourier space, using fast Fourier transforms;
  \item \label{Cov_steps3} Calculate isotropic spatial power spectra $\bar P_{\mu\nu}(k)$ in (\ref{ps_den_final})-(\ref{ps_dm_final}) and find appropriate fitting functions;
  \item \label{Cov_steps4} Obtain spatial correlation functions $\zeta_{\mu\nu}(r)$ in (\ref{comb_cf}) by reverse Fourier transforms.
\end{enumerate}
Details are given in Appendix \ref{pip_CM}.

\subsubsection{Momentum power spectra}

\begin{figure}
  \centering
  \includegraphics[width=0.47\hsize]{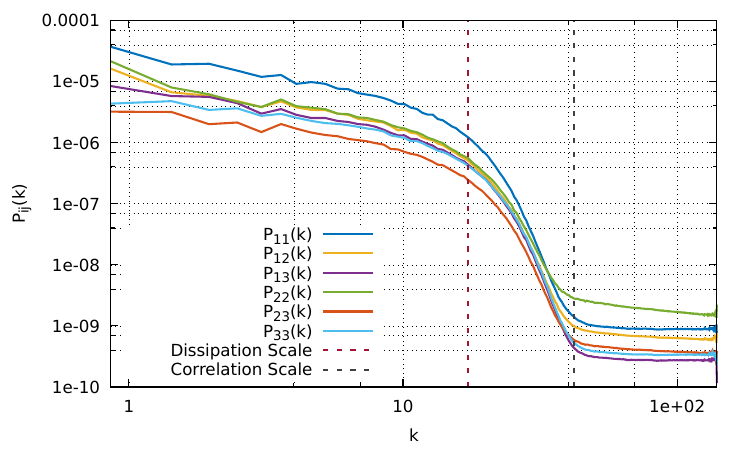}\hfill
  \includegraphics[width=0.47\hsize]{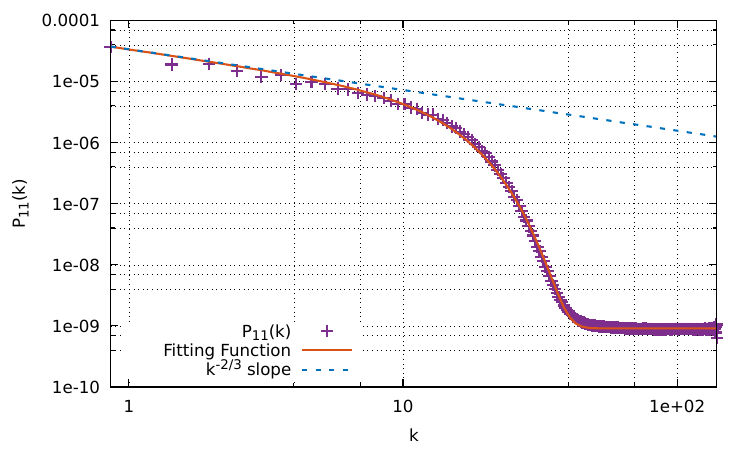}
\caption{\emph{Left}: The six independent components of the momentum power spectrum are shown here. The $k$ axis is scaled by $\frac{2\pi}{L}$ here and below. The red vertical line at $k=17.5$ indicates the dissipation scale $l_d$, and the grey vertical line at $k=42$ indicates the correlation scale $l_c$. \emph{Right}: Momentum power spectrum $P_{11}$ adapted from the simulation (purple dots) together with an individual fit function of the form (\ref{Allmom_eq2}, orange line). The blue dashed line illustrates the $k^{-2/3}$ slope for reference.}
\label{fig:AllMom}
\end{figure}

The momentum power spectra $P_{ij}(k)$ obtained this way are shown in Fig.~\ref{fig:AllMom}. They are all of similar shape, characterised by a first (larger) scale at $k \approx 17.5$ and a second (smaller) scale at $k \approx 42$. These two scales, which are roughly determined by eyes, represent respectively the dissipation scale $l_d$ and the correlation scale $l_c$. On scales smaller than $l_c$, the particles are significantly less correlated due to their random thermal motion. Moreover, since the simulated density field is resolved into a finite number of particles, shot noise becomes visible at small scales, preventing the power spectra from falling to zero. On scales larger than $l_d$, the curves approach power laws indicating turbulence. In between these two scales, the dissipation process becomes visible, where the turbulent kinetic energy is converted to heat.

The shape of the isotropic momentum power spectra are well described by the function
\begin{equation}\label{Allmom_eq2}
  P_{ij}(k) = c_{1ij}+\frac{c_{2ij}}{k^{\alpha_{ij}}}
  \exp\left(-\frac{k^2}{2\sigma_{ij}^2}\right)
\end{equation}
with fit parameters $c_{1ij}$, $c_{2ij}$ and $\sigma_{ij}$. The exponents $\alpha_{ij}$ are close enough to $2/3$ for keeping them fixed at this value. Best-fitting values for the remaining coefficients are listed in Tab.~\ref{tab_mom_ps}, and one individual fit is shown in the right panel of Fig.~\ref{fig:AllMom}.

\begin{table}
\caption{Best-fitting parameters for all independent momentum power spectra. The small but non-zero first value for all curves reflects the shot-noise level.}
  \centering
  \begin{tabular}{|c|c|c|c|}
    \hline
    & $c_{1ij}\cdot10^{-10}$ & $c_{2ij}\cdot10^{-6}$ & $\sigma_{ij}$ \\
    \hline
    $P_{11}$ & $9.12$ &$33.8$ & $9.74$ \\
    $P_{12}$ & $6.62$ &$12.3$ & $10.12$ \\
    $P_{13}$ & $2.72$ &$9.98$ & $10.37$\\
    $P_{22}$ & $18.3$ &$12.2$ & $10.56$\\
    $P_{23}$ & $3.93$ &$5.18$ & $10.49$\\
    $P_{33}$ & $3.42$ &$8.51$ & $10.37$\\
    \hline
  \end{tabular}
\label{tab_mom_ps}
\end{table}

Even though the data become noisier towards small wave numbers, the fit function follows the data accurately, and the common exponent $\alpha_{ij} = 2/3$ well describes the slope of the power spectra at large scales.

\subsubsection{Density power spectrum}

\begin{figure}
  \includegraphics[width=0.49\hsize]{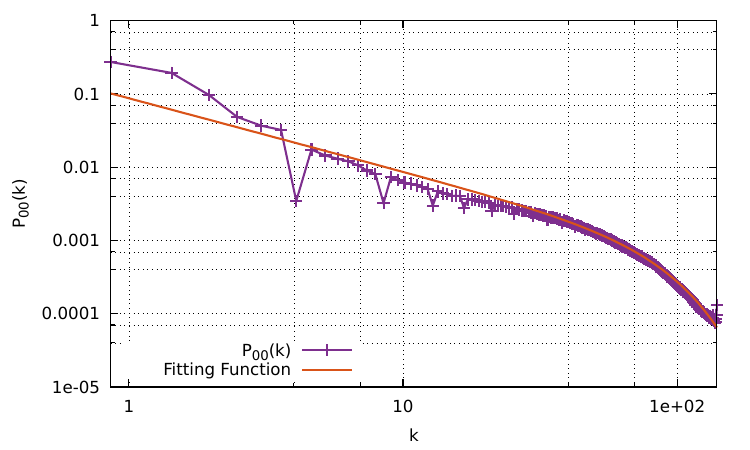}\hfill
  \includegraphics[width=0.49\hsize]{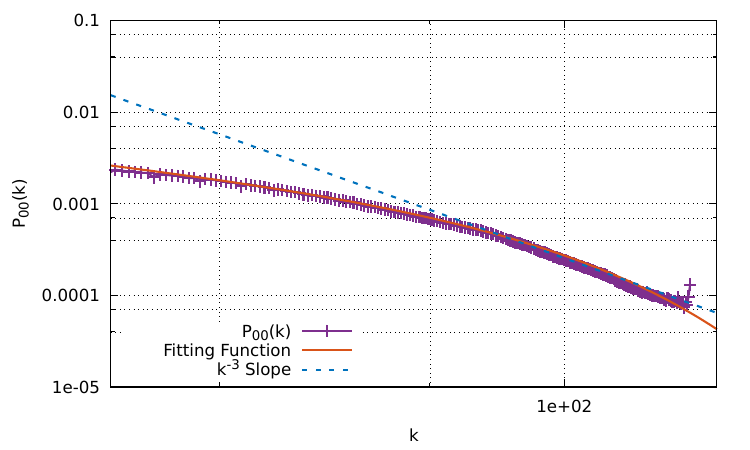}
\caption{Density power spectra extracted from the simulation (purple points) and modelled by the fit function (\ref{density_fit_func}, orange line). \emph{Left}: complete $k$ range allowed by the simulation data. \emph{Right}: small scales, i.e.\ large wave numbers. The blue dashed line represents the $k^{-3}$ slope.}
\label{fig:dd}
\end{figure}

The density power spectrum $P_{00}(k)$ shown in Fig.~\ref{fig:dd} suggests a fit function of the form
\begin{equation}\label{density_fit_func}
  P_{00}(k) = \frac{c_{200}}{k}
  \exp\left(-\frac{k^2}{2\sigma_{00}^2}\right)\;,
\end{equation}
with best-fitting values for the parameters $c_{100}$, $c_{200}$ and $\sigma_{00}$ given in Tab.~\ref{tab_den_ps}.

\begin{table}
\caption{Best-fitting parameters for the isotropic density spatial power spectrum $P_{00}(k)$}
  \centering
  \begin{tabular}{|c|c|c|}
    \hline
    &  $c_{200}$ & $\sigma_{00}$ \\
    \hline
    $P_{00}$ & $8.70\cdot10^{-2}$ & $65.64$ \\
    \hline
  \end{tabular}
\label{tab_den_ps}
\end{table}

Unlike the momentum power spectra shown in Fig.~\ref{fig:AllMom}, the density power spectrum keeps decreasing for increasing wave numbers $k$. At small wave numbers, the data become noisy, but the fit function still follows the data points reasonably well. Since we do not want to describe one individual simulation precisely, but to extract appropriate model functions from it, we deem our fit function (\ref{density_fit_func}) acceptable.

The right panel of Fig.~\ref{fig:dd} enlarges the density power spectrum at smaller scales. Notice that the power spectrum changes shape at around $k = 85$. At smaller scales, i.e.\ larger wave numbers, the power spectrum develops a $k^{-3}$ slope, indicating structure formation at small scales.

\begin{figure}
  \centering
  \includegraphics[width=0.47\hsize]{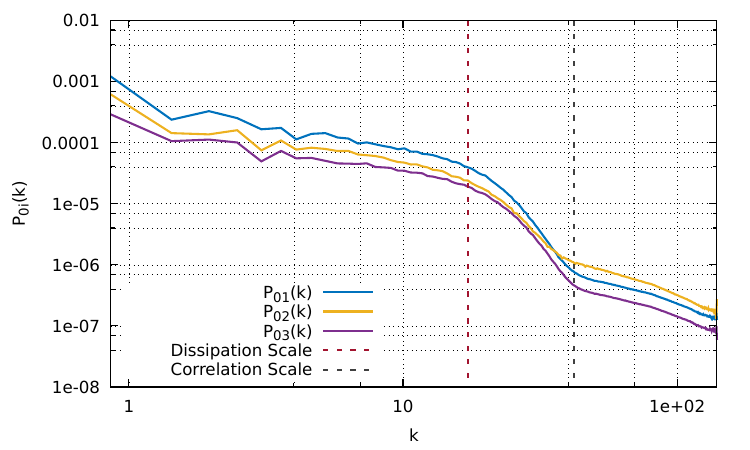}\hfill
  \includegraphics[width=0.47\hsize]{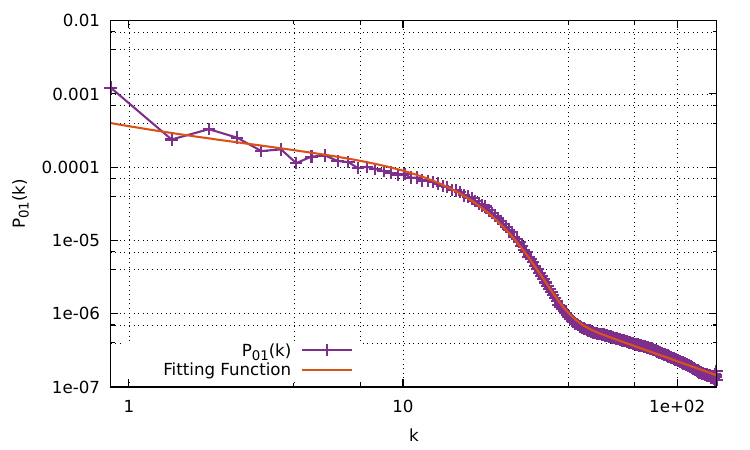}
\caption{\emph{Left}: Cross power spectrum between the density and the momentum components. Similar to the momentum power spectra shown in Fig.~\ref{fig:AllMom}, the orange vertical line at $k=17.5$ represents the dissipation scale $l_d$, and the yellow vertical line at $k=42$ the correlation scale $l_c$. \emph{Right}: Density-momentum cross power spectrum $P_{01}$, overlaid with the fit function (\ref{den-mom-eq}), with exponents $\beta_{0i} = 4/3$ and $\alpha_{0i} = 1/3$ for all $i$. The purple dots are the data points, the green lines show the fits with parameters listed in Tab.~\ref{tab_denmom_ps}.}
\label{fig:all_den_mom}
\end{figure}

\subsubsection{Momentum-density cross-power spectrum}

Next, we turn to the cross power spectrum between density and momenta, illustrated in Fig.~\ref{fig:all_den_mom}. The shapes of these three cross power spectra closely resemble that of the momentum and density spectra shown in Figs.~\ref{fig:AllMom} and \ref{fig:dd}, with similar parameter values. A suitable fit function is
\begin{equation}\label{den-mom-eq}
  P_{0i}(k) = \frac{c_{10i}}{k^{\beta_{0i}}}+
  \frac{c_{20i}}{k^{\alpha_{0i}}}\exp\left(
    -\frac{k^2}{2\sigma_{0i}^2}
  \right)\;.
\end{equation}
The additional exponent $\beta_{0i}$ is introduced to follow the small-scale tails in Fig.~\ref{fig:all_den_mom}. Fitting this function to the simulation data, we find equal exponents $\alpha_{0i} = 1/3$ and $\beta_{0i} = 4/3$ for all directions $i$. The remaining fit parameters are listed in Tab.~\ref{tab_denmom_ps}. As the right panel of Fig.~\ref{fig:all_den_mom} shows, the fit function follows the data points very closely.

\begin{table}
\caption{Fit parameters for all density-momentum cross power spectra}
  \centering
  \begin{tabular}{|c|c|c|c|}
    \hline
    & $c_{10i}\cdot10^{-4}$ & $c_{20i}\cdot10^{-4}$ & $\sigma_{0i}$ \\
    \hline
    $P_{01}$ & $1.04$ & $2.58$ & 11.93 \\
    $P_{02}$ & $1.47$ & $1.28$ & 12.28 \\
    $P_{03}$ & $0.61$ & $1.03$ & 12.73 \\
    \hline
  \end{tabular}
\label{tab_denmom_ps}
\end{table}

\subsubsection{Correlation functions and covariance matrix}

\begin{figure}
  \includegraphics[width=0.32\hsize]{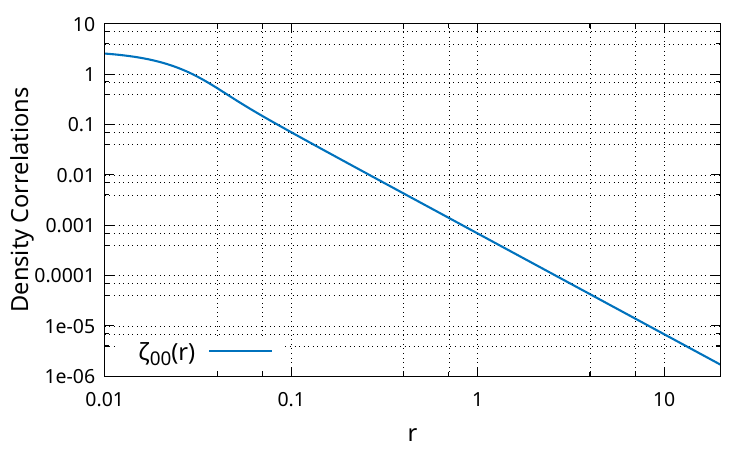}\hfill
  \includegraphics[width=0.32\hsize]{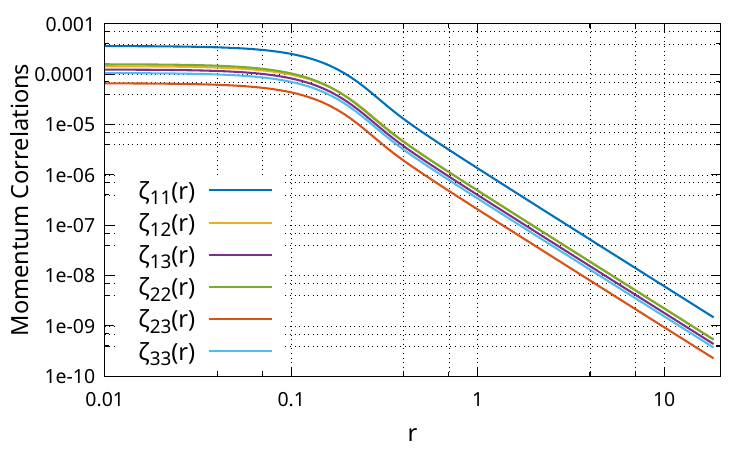}\hfill
  \includegraphics[width=0.32\hsize]{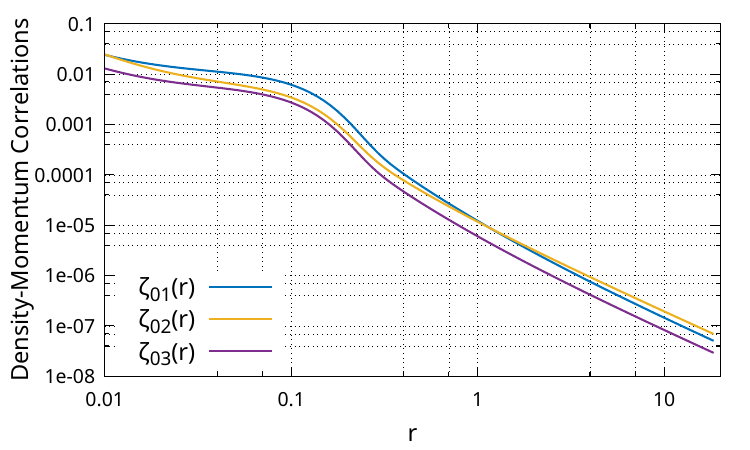}
\caption{The density, momentum, and density-momentum correlation functions of (\ref{combined_cf_fit}) are shown here in the left, centre, and right panels, respectively. The horizontal axes are scaled by $L=0.1H$, i.e.\ the same length scale as the simulation boxes.}
\label{fig:cf_all}
\end{figure}

From the density, momentum, and density-momentum power spectra deduced and modelled above, we can now calculate the correlation functions $\zeta_{\mu\nu}(r)$ via the Fourier transform (\ref{comb_cf}), thereby dropping delta distributions originating from shot-noise terms. The results are
\begin{equation}\label{combined_cf_fit}
  \zeta_{\mu\nu}(r) =
  \begin{cases}
    \displaystyle
    \frac{1}{\sqrt{2}\pi^2}\frac{c_{200}\sigma_{00}}{r}\,F\left(
      \frac{\sigma_{00}}{\sqrt{2}}r
    \right) & \mbox{for} \quad \mu,\nu=0 \\[10pt]
    \displaystyle
	  \frac{1}{2^{5/6}\pi^2}c_{2ij}\sigma_{ij}^{7/3}\Gamma\left(
	    \frac{7}{6}
	  \right){_1F_1}\left(
	    \frac{7}{6},\frac{3}{2},-\frac{1}{2}\sigma_{ij}^2r^2
	  \right) & \mbox{for} \quad \mu,\nu=1,2,3 \\[10pt]
    \displaystyle
	  \frac{\sqrt{3}c_{10i}}{4\pi^2r^{5/3}}\Gamma\left(\frac{2}{3}\right)+
	  \frac{1}{2^{2/3}\pi^2}c_{20i}\sigma_{0i}^{8/3}
	  \Gamma\left(\frac{4}{3}\right){_1F_1}\left(
	    \frac{4}{3},\frac{3}{2},-\frac{1}{2}\sigma_{0i}^2r^2
	  \right) & \mbox{for} \quad \mu=0\;,\; \nu=1,2,3
  \end{cases}\;.
\end{equation}
Here, $F$ is the Dawson integral, and ${_1F_1}[z]$ is the Kummer confluent hypergeometric function. These correlation functions are shown in Fig.~\ref{fig:cf_all}. We note the similarity of the curves in its three panels. In particular, they all change shape at a relatively small scale $r_d$, above which they fall more steeply.

For any pair $(i,j)$ of particles, the complete correlation matrix can now be written in the form
\begin{equation}\label{sym_cij}
  \bar C_{ij}(r_{ij}) =
  \begin{pmatrix}
    \zeta_{00}(r_{ij}) & \zeta_{0k}^\top(r_{ij}) \\
    \zeta_{0k}(r_{ij}) & \zeta_{kl}(r_{ij})
  \end{pmatrix}
\end{equation}
with $k,l\in1,2,3$ and $r_{ij}=|\vec{q}_i-\vec{q}_j|$. The covariance matrix for the entire particle ensemble defined in (\ref{cov_matr}) is then $\bar{C}=\bar{C}_{ij}\otimes E_{ij}$.

Finally, we isotropize the momentum covariance matrix. Since the spatial scales we are considering are small compared to the scale height $H$ of the protoplanetary disk, we can average the modified momentum covariance matrix $\Bar{C}_{p_ip_j}$ to read
\begin{equation}
  \Bar{C}_{p_ip_j} =
  \begin{pmatrix}
    b_0 & b_1 & b_1 \\
    b_1 & b_0 & b_1 \\
    b_1 & b_1 & b_0 \\
  \end{pmatrix}\,\zeta_{pp}(r_{ij})
  \label{mod_mom_cov}
\end{equation}
with components calculated from Table \ref{tab_mom_ps} as
\begin{equation}
  b_0 = \frac{1}{3}\left(c_{211}+c_{222}+c_{233}\right)\;,\quad
  b_1 = \frac{1}{3}\left(c_{212}+c_{213}+c_{223}\right)
\label{eq:2}
\end{equation}
and the averaged momentum correlation function
\begin{equation}
  \zeta_{pp}(r) = \frac{1}{2^{5/6}\pi^2}\bar{\sigma}^{7/3}\Gamma\left(
    \frac{7}{6}
  \right){_1F_1}\left(
    \frac{7}{6},\frac{3}{2},-\frac{1}{2}\bar{\sigma}^2r^2
  \right)
\end{equation}
where the averaged dispersion is
\begin{equation}\label{sigmabar}
  \bar{\sigma} = \frac{1}{6}\sum_{j\ge i=1}^3\sigma_{ij}\;.
\end{equation}

For calculating non-linear density power spectra with KFT, we now rotate from the coordinate frame of the simulation box into a coordinate frame in which the relative position vector $\vec r_{ij} = \vec q_i-\vec q_j$ between the particles is the polar axis. We first define the projectors
\begin{equation}\label{def_projects}
  \pi_\parallel = \hat q\otimes\hat q\;\quad
  \pi_\perp = \mathbb{1}_{3}-\pi_\parallel
\end{equation}
parallel and perpendicular to $\hat q = \vec r_{ij}/|\vec r_{ij}|$. There, $\bar C_{p_ip_j}$ is of the form
\begin{equation}
  \bar C_{p_ip_j} = -m_1\mathbb{1}_{3}-m_2\hat q\otimes\hat q.
\label{eq:5}
\end{equation}
As we know, rotating any matrix, its trace and determinant must remain unchanged. The eigenvalues of (\ref{mod_mom_cov}) are
\begin{equation}
 \lambda_{1,2} = (b_0-b_1)\zeta_{pp}(r_{ij}) , \quad \lambda_3 = (b_0+2b_1)\zeta_{pp}(r_{ij}),
  \label{eq:3}
\end{equation}
thus its trace and determinant are
\begin{equation}
  \text{tr}\Bar{C}_{p_ip_j}  = 3b_0\zeta_{pp}(r_{ij}) ,\quad \det\Bar{C}_{p_ip_j}  = (b_0-b_1)^2(b_0+2b_1)\zeta_{pp}^3(r_{ij})\;.
  \label{eq:4}
\end{equation} 
On the other hand, the eigenvalues of the momenta covariance matrix in (\ref{eq:5}) are
\begin{equation}
 l_{1,2} = -m_1\;,\quad l_3 = -(m_1+m_2)\;.
  \label{eq:6}
\end{equation} 
Identifying these with the eigenvalues in (\ref{eq:3}), we have the correlation functions
\begin{equation}
  m_1 (r_{ij}) = (b_1-b_0)\zeta_{pp}(r_{ij})\;,\quad m_2 (r_{ij})= -3b_1\zeta_{pp}(r_{ij})\;,
  \label{eq:7}
\end{equation} 
which ensures that trace and determinant of the same matrix will be identical. Therefore, by setting $m_1$ and $m_2$ as above, we are able to rotate our momenta covariance matrix $\Bar{C}_{p_ip_j}$ in the $\hat{q}$ space.

\section{Non-Linear Density Power Spectrum}\label{sec:non_ps}

In this section, we will first define trajectories for freely streaming particles, then calculate the evolved, non-linear density power spectrum within KFT.

\subsection{Freely Streaming Particles}

Freely streaming particles are defined by the absence of interactions between them. Their equation of motion is Hamilton's equation,
\begin{equation}
  \Dot{x}-\mathcal{J}\nabla_xH=0
\end{equation}
where $x:=\{\vec{q},\vec{p}\}^\top$ is the phase-space position, $H=\vec p^2/(2m)$ is the free Hamilton function, and
\begin{equation}
  \mathcal{J} =
  \begin{pmatrix}
    0 & \mathbb{1}_3 \\
    -\mathbb{1}_3 & 0
  \end{pmatrix}
\end{equation}
is the symplectic matrix, with $\mathbb{1}_n$ denoting the unit matrix in $n$ dimensions. The particle trajectories are thus
\begin{equation}\label{eq:par_traj}
  \bar{x}(t) = G(t,0)x^{(i)}
\end{equation}
beginning at the initial phase-space position $x^{(i)}:=\{\vec{q}^{(i)},\vec{p}^{(i)}\}$. The Green’s function $G$ is a $6\times6$ matrix. With $g_{qq}(t,t') = g_{pp}(t,t') = 1$ and $g_{qp}(t,t') = (t-t')/m$,
\begin{equation}\label{gf}
  G(t,t') =
  \begin{pmatrix}
    \mathbb{1}_3 & (t-t')m^{-1}\mathbb{1}_3 \\
    0 & \mathbb{1}_3
  \end{pmatrix}\;.
\end{equation}
For simplicity, we abbreviate $g_{qp}(t,0):=g_{qp}(t)$ and $m=1$.

\subsection{Non-Linear Density Power Spectrum}\label{sec:non_lin_dps}

In this section, we derive the non-linear density power spectrum for dust particles. Following an asymptotic analysis for this expression, we find that the density power spectrum will necessarily develop a universal $k^{-3}$ tail for large $k$, suggesting that structures form below certain length scales already at very early times in the evolution of protoplanetary disks.

\subsubsection{Factorisation of the Free Generating Functional}

Power spectra are derived in KFT by applying density operators to the generating functional. The one-particle density operator in Fourier space is given by 
\begin{equation}\label{eq2.26}
    \hat{\rho}_j(1):=\exp\left(-\Vec{k}_1\cdot\frac{\delta}{\delta \Vec{J}_{q_j}(t_1)}\right),
\end{equation}
where we have introduced the conventional short-hand notation $(k_s,t_s):=(s)$, see more detailed derivations in \cite{bartelmann2019cosmic}. For indistinguishable particles, which we shall henceforth assume, we can set the particle index $j$ to an arbitrary value without loss of generality. Since the operator in (\ref{eq2.26}) is an exponential of a derivative with respect to the generator field, it corresponds to a finite translation of the generator field. After applying $r\geq 1$ of these operators, the generator field is translated by
  
  \begin{gather}\label{eq2.27}
     \tens{J}\to \tens{J}-\sum^{r}_{j=1}\delta(t'-t_j)\Vec{k}_j\cdot
      \begin{pmatrix}
      1\\
      0
      \end{pmatrix}
      \otimes \Vec{e}_j.
  \end{gather}

Since we are studying the power spectrum of freely moving dust particles here, we use the free generating functional $Z_0$. After applying two density operators and setting the source field $\boldsymbol J$ to zero, it acquires the form
\begin{equation}\label{z_0}
  Z_0[\tens L] =\hat\rho_1(1)\hat\rho_2(2) Z_0[\boldsymbol{J}]\big|_{\boldsymbol J=0}= \int\D\tens q\D\tens p\,
  P(\tens q,\tens p)\,
  \exp\left(
    \I\left<\tens L_q,\tens q\right>+
    \I\left<\tens L_p,\tens p\right>
  \right)
\end{equation}
with the components
\begin{equation}
  \tens L_q = -\sum_{j=1}^{2}\Vec{k}_j\otimes e_j \;,\quad
  \tens L_p = -\sum_{j=1}^{2}\Vec{k}_jg_{qp}(t_j)\otimes e_j
\end{equation}
of the translation tensor generated by the density operators. In (\ref{z_0}), $\tens q$ and $\tens p$ are the initial particle positions and momenta, combined as $(\tens q,\tens p) := \tens x^{(i)}$. The probability distribution $P(\tens q,\tens p)$ was given in (\ref{final_prob_dist}).

Since the correlation operator $\mathcal{C}(\tens p)$ in $P(\tens q,\tens p)$ has the hierarchical structure given in (\ref{eq:ctp}), we can drop part of its terms as negligible according to the following relations
\begin{equation}
  \Bar{C}_{\bar{\rho}_j\bar{\rho}_j} > \Bar{C}_{\bar{\rho}_jp_k} \gg
  \Bar{C}_{\bar{\rho}_jp_k}^2 \;,\quad
  \Bar{C}_{\bar{\rho}_j\bar{\rho}_j} = \zeta_{00}(0) >
  \Bar{C}_{\bar{\rho}_j\bar{\rho}_k} = \zeta_{00}(r_{jk} > 0)\;,\quad \big|t_{p_k}\big|\leq 5.81\cdot 10^{-3}\;,
\end{equation}
with indices $j$ and $k$ indicating an arbitrary pair of distinct particles and $\zeta_{00}(r)$ given in (\ref{combined_cf_fit}). We can thus ignore the terms including $\Bar{C}_{\bar{\rho}_jp_k}t_{p_k}$ and $\Bar{C}_{\bar{\rho}_j\bar{\rho}_k}$ in (\ref{eq:ctp}) and approximate the correlation operator $\mathcal{C}(\tens p)$ by the constant
\begin{equation}
  \mathcal{C}(\tens p) = \mathcal{C} =
  \left(\zeta_{00}(0)+\mu_{\bar{\rho}}^2\right)^N
\end{equation}
with the mean value of the density field $\Bar{\rho}$ given by $\mu_{\Bar{\rho}}=\sqrt{a_1}a_2$ as shown in (\ref{inverse_func_rho'}).

We can then integrate over the momenta $\tens p$ in (\ref{z_0}) to obtain
\begin{equation}
  Z_0[\tens L] = \mathcal{C}N^{-N}\int\D\tens q\exp\left(
    -\frac{1}{2}\tens L_{p}^\top\Bar{C}_{pp}\tens L_{p}
  \right)\exp\left(
  \I\langle\tens L_q,\tens q\rangle
  \right)
\end{equation}
where
\begin{equation}
  \langle\tens L_q,\tens q\rangle =
  -\Vec{k}_1\cdot\Vec{q}_1-\Vec{k}_2\cdot\Vec{q}_2\;.
\end{equation}

With the integration over all other particle positions $\vec q_i$ with $i=3,..,N$ being
\begin{equation}
  \int d q_3... d q_N=V^{N-2}
\end{equation}
where the factor $V\equiv\int_{q_i}$ is the volume of the 3-dimensional space, we introduce $\vec r = \vec q_2-\vec q_1$, integrate over $\vec q_1$, and arrive at
\begin{equation}\label{Z-final}
  Z_0[\tens L] = \mathcal{C}V^{N-2}N^{-N}(2\pi)^3
  \delta_\mathrm{D}\left(\Vec{k}_1+\Vec{k}_2\right)
  \mathcal{P}_{0}(k_1,t)
\end{equation}
with the non-linearly evolved, free density-fluctuation power spectrum
\begin{equation}\label{ori_full_ps}
  \mathcal{P}_{0}(k_1,t) = \int_r \E^{-g_{qp}^2(t)k_1^2\left(
    (m_1(r)-m_1(0))+\mu^2\left(
      m_2(r)-m_2(0)
    \right)
  \right)}\E^{-\I k_1r\mu}\;,
\end{equation}
where $\mu$ is the cosine of the angle enclosed by $\vec k$ and $\vec r$. The correlation functions $m_1(r)$ and $m_2(r)$ are given in (\ref{eq:7}). We have used here that $\vec{k}_1+\vec{k}_2=0$ as required by the delta distribution in (\ref{Z-final}).

\subsubsection{Linear density power spectrum from KFT}
\label{Den_ps}

Before we evaluate the non-linear, free density-fluctuation power spectrum given in (\ref{ori_full_ps}), we have a closer look at the correlation functions $m_1(r)$ and $m_2(r)$. Instead of the full expression for $\zeta_{pp}(r)$, we introduce a simpler approximation reproducing its asymptotic behaviour for $r\to0$ and $r\to\infty$. We define $\zeta(r)$ by
\begin{equation}\label{zeta}
  \zeta_{pp}(r) \sim \zeta(r) = \frac{0.04}{(0.0075+r^2)^{7/6}}
\end{equation}
as well as $a_1(r)$ and $a_2(r)$ by
\begin{equation}\label{a1}
  |m_1(r)| = -m_1(r) \sim a_1(r) = (b_0-b_1)\zeta(r) =
  \frac{3.5\cdot 10^{-7}}{(0.0075+r^2)^{7/6}}
\end{equation}
and
\begin{equation}\label{a2}
  |m_2(r)| = -m_2(r) \sim a_2(r) = 3b_1\zeta(r) =
  \frac{1.1\cdot 10^{-6}}{(0.0075+r^2)^{7/6}}
\end{equation}
with $b_0$ and $b_1$ from (\ref{eq:2}). For both small and large $r$, the asymptotic behaviour of the approximation $\zeta$ agrees with that of the original functions $\zeta_{pp}(r)$, $|m_1(r)|$ and $|m_2(r)|$ very well. Furthermore, at large scales, the fit functions fall off $\propto r^{-7/3}$ like the original functions.

We split $\mathcal{P}_{0}(k_1,t)$ from (\ref{ori_full_ps}) into
\begin{equation}\label{eq:new_non_p}
  \mathcal{P}_{0}(k_1,t) = \int_r\left(
    \E^{-g_{qp}^2(t)k_1^2\left(
      m_1(r)+\mu^2 m_2(r)
    \right)}-1
  \right)\E^{y-\I k_1r\mu}+\int_r\E^{y-\I k_1r\mu}
\end{equation}
with
\begin{equation}\label{y(t,mu)}
  y(t,\mu) = g_{qp}^2(t)k_1^2(m_1(0)+\mu^2 m_2(0))
\end{equation}
and denote the first term on the right-hand side as $\mathcal{P}_\mathrm{free}(k_1,t)$,
\begin{align}
  \mathcal{P}_\mathrm{free}(k_1,t) &= 2\pi\int_0^\infty\D r\,r^2l(r)
  \nonumber\\ &=
  2\pi\int_0^\infty\D r\,r^2\int_{-1}^{1}\D\mu\left(
    \E^{-g_{qp}^2(t)k_1^2\left(
      m_1(r)+\mu^2 m_2(r)
    \right)}-1
  \right)\E^{y-ik_1r\mu}\;.
\label{eq:71}
\end{align}
At finite time $t$ and wave number $k_1$, the expression in parentheses satisfies
\begin{equation}
  \lim_{r\to\infty}\left(
    \E^{-g_{qp}^2(t)k_1^2\left(
      m_1(r)+\mu^2 m_2(r)
    \right)}-1
  \right) =
  \lim_{r\to \infty}g_{qp}^2(t)k_1^2(b_0-b_1+3b_1\mu^2)
  \frac{0.04}{r^{7/3}} = 0\;.
\end{equation}

In the limit of $r\to \infty$, the integrand $r^2l(r)$ of (\ref{eq:71}) becomes
\begin{align}\label{r2lr_asym}
  \lim_{r\to\infty}r^2l(r) &=
  \lim_{r\to\infty}0.04\,g_{qp}^2(t)k_1^2\,\frac{r^2}{r^{7/3}}
  \int_{-1}^{1}\D\mu(b_0-b_1+3b_1\mu^2)
  \E^{y(t,\mu)k_1^2}\cos(k_1 r \mu) \nonumber\\ &<
  0.08\,g_{qp}^2(t)k_1^2(b_0+2b_1)\E^{y(t,0)k_1^2}
  \lim_{r\to\infty}\frac{1}{r^{1/3}}\frac{\sin(k_1 r)}{k_1 r} = 0\;,
\end{align}
which guarantees that $\mathcal{P}_\mathrm{free}(k_1,t)$ converges at large $r$. We have used here that $y(t,\mu)$ is a negative, monotonically decreasing function of $\mu^2$ with a maximum at $y(t,0)$. The more detailed discussion in Appendix \ref{app3:more_ps} shows that we can safely ignore the second integral on the right-hand side of (\ref{eq:new_non_p}) and write down
\begin{equation}\label{final_full_ps}
  \mathcal{P}_\mathrm{free}(k_1,t) =
  \int_r\left(
    \E^{-g_{qp}^2(t)k_1^2\left(
      m_1(r)+\mu^2 m_2(r)
    \right)}-1
  \right)\E^{y-ik_1r\mu}
\end{equation}
for the final expression for the non-linearly evolved density power spectrum.

At very early times, $g_{qp}\to 0$, the exponential in (\ref{final_full_ps}) can be Taylor approximated to first order,
\begin{equation}\label{taylor_P}
  \mathcal{P}_\mathrm{free}(k_1,t) \approx
  \mathcal{P}_\mathrm{lin}(k_1,t) = -g_{qp}^2(t)k_1^2\int_r\left(
    m_1(r)+\mu^2 m_2(r)
  \right)\E^{-\I k_1r\mu}\;.
\end{equation}
Integrating over $\mu$ and $\varphi$ gives
\begin{equation}\label{taylor_P_final}
  \mathcal{P}_\mathrm{lin}(k_1,t) =
  -4\pi g_{qp}^2(t)\int_0^\infty\D r\,\Bigg[
    \left(
      \frac{6b_1\sin(k_1r)}{k_1r}-6b_1\cos(k_1r)
    \right)\zeta_{pp}(r)-
    \underbrace{(b_0+2b_1)k_1r\sin(k_1r)\zeta_{pp}(r)}_{(1)}
  \Bigg]\;.
\end{equation}
The integral $(1)$ can be carried out to give
\begin{equation}
  4\pi g_{qp}^2(t) \int_0^\infty\D r\,
  (b_0+2b_1)k_1r\sin(k_1r)\zeta_{pp}(r) =
  g_{qp}^2k_1^2(b_0+2b_1)\bar{P}(k_1)
\end{equation}
with the averaged initial momentum power spectrum
\begin{equation}\label{eq:ave_imps}
  \bar{P}(k_1) = \int_r\zeta_{pp}(r)\E^{-\I\vec{k}_1\cdot\vec{r}} =
  \frac{1}{k_1^{2/3}}\exp\left(
    -\frac{k_1^2}{2\bar{\sigma}^2}
  \right)\;.
\end{equation}
Since $\bar{P}(k)\propto k^{-2/3}$ at small wave numbers, i.e.\ at large scales, the integral $(1)$ is proportional to $k_1^{4/3}$ when $k_1$ is small. On the other hand, substituting $r\to r_1/k_1$ in (\ref{taylor_P_final}) gives
\begin{equation}\label{tayP_nr}
  \mathcal{P}_\mathrm{lin}(k_1,t) =
  -\frac{4\pi g_{qp}^2(t)}{k_1}\int_0^{\infty}\D r_1\left(
    \frac{6b_1\sin(r_1)}{r_1}-6b_1\cos(r_1)-(b_0+2b_1)r_1\sin(r_1)
  \right)\zeta_{pp}\left(\frac{r_1}{k_1}\right)\;.
\end{equation}

When $k_1\to 0$, according to (\ref{zeta}), we have $\zeta_{pp}\left(r_1/k_1\right)\propto \left(k_1/r_1\right)^{7/3}$, thus the linear density power spectrum satisfies $\mathcal{P}_\mathrm{lin}(k_1,t)\propto k_1^{4/3}$, which coincides with the conclusion on the integral over $(1)$. For $k_1\to\infty$, the result of the integral in (\ref{tayP_nr}) becomes independent of $k_1$, thus the density power spectrum satisfies $\mathcal{P}_\mathrm{lin}(k_1,t)\propto k_1^{-1}$ for small scales.

The linear density power spectrum $\mathcal{P}_\mathrm{lin}(k_1,t)$ defined in (\ref{taylor_P_final}) is shown in Fig.~\ref{fig:ps_com} at $t=4 \pi$ together with its asymptotic behaviour at both small and large $k_1$, which verifies our asymptotic analysis for both small and large scales. Note that the radical change of the curve around scale $k_1\approx 10$ is due to a sign shift of computing (\ref{taylor_P_final}). Though the power spectrum is expected to be positive, the integral over the linear Taylor expansion of the modified density power spectrum (\ref{final_full_ps}) could as well become negative.

\subsubsection{Non-linear density power spectrum from KFT}

We now return to the full non-linear density power spectrum (\ref{final_full_ps}) and integrate over the angular coordinates $\varphi$ and $\mu$ to write
\begin{equation}
  \mathcal{P}_\mathrm{free}(k_1,t) = 2\pi\int_0^\infty\D r\,r^2l(r)
\end{equation}
as in (\ref{eq:71}). The function $l(r)$ can be expressed in a closed, but somewhat opaque form. For large wave numbers, the radial integrand $r^2l(r)$ develops a sharp peak at $r = 0$, closely resembling a delta distribution. For large $r$, it falls off proportional to $r^{-4/3}$, confirming that our results for the non-linear density power spectrum $\mathcal{P}_\mathrm{free}(k_1,t)$ as shown in (\ref{r2lr_asym}) converge.

\begin{figure}
  \centering
  \includegraphics[width=0.47\hsize]{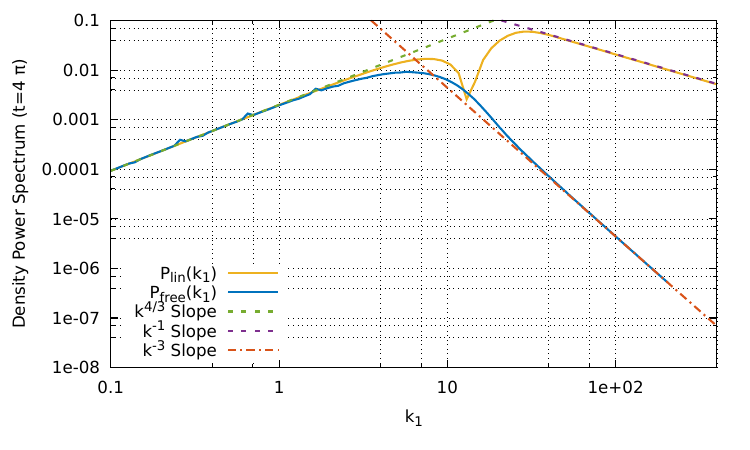}\hfill
  \includegraphics[width=0.47\hsize]{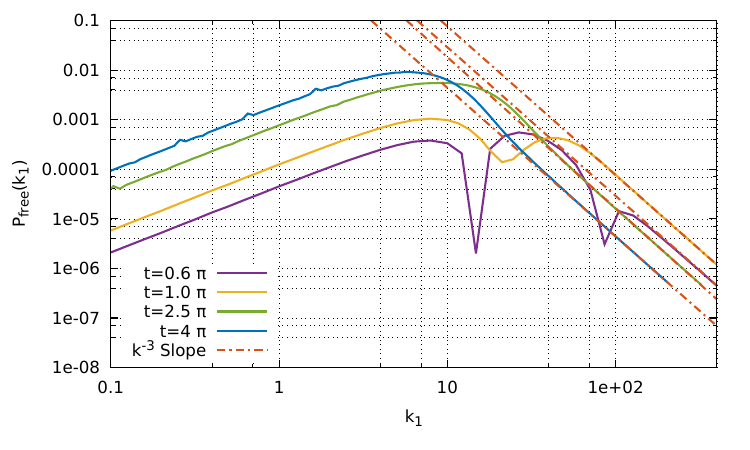}
\caption{\emph{Left}: The linearly evolved density power spectrum $\mathcal{P}_\mathrm{lin}(k_1,t)$ (blue line) compared to the free non-linear density power spectrum $\mathcal{P}_\mathrm{free}(k_1,t)$ (yellow line). The asymptotic behaviour at large (green dashed line) and small scales is also indicated. At small scales, the linear power spectrum falls off like $k^{-1}$ (purple dashed line), while the free non-linear power spectrum develops a $k^{-3}$ slope (orange dash-dotted line). \emph{Right}: Non-linear density power spectra $\mathcal{P}_\mathrm{free}(k_1,t)$ together with their small scale asymptotic behaviour (orange dash-dotted lines) at the four different times $t=0.6\pi$ (purple line), $t=1\pi$  (yellow line), $t=2.5\pi$ (green line) and $t=4\pi$ (blue line).}
\label{fig:ps_com}
\end{figure}

Figure \ref{fig:ps_com} shows the directly integrated function $\mathcal{P}_\mathrm{free}(k_1,t)$ and compares it to the density power spectrum $\mathcal{P}_\mathrm{lin}(k_1,t)$ linearly evolved to $t=4\pi$. At large scales, it reproduces the $k^{4/3}$ slope of $\mathcal{P}_\mathrm{lin}(k_1,t)$. At small scales, however, nonlinear evolution causes the spectrum to fall off like $k^{-3}$. This asymptotic slope sets in near $k\approx 25$.

\subsubsection{Small-scale asymptotic behaviour}

To better understand the $k^{-3}$ behaviour at small scales, we adopt an asymptotic analysis developed by \cite{sara}. Consider integrals of the form
\begin{equation}
  P(k) = \int_D\E^{-|k|^sf(x)}g(x)\E^{\I\vec{k}\cdot \vec{x}}\D x\;,
  \quad s\geq 2
\label{eq:81}
\end{equation}
over a possibly unbounded domain $D\subset\mathbb{R}^3$. Following \cite{sara}, if the function $f(x)$ satisfies the three criteria that
\begin{enumerate}
  \item \label{C1} $f(x)$ has an isolated global minimum at $\vec{x}=0$;
  \item \label{C2} $f(x)$ is quadratically integrable on $\mathbb{R}^3$; and
  \item \label{C3} the Hessian matrix $A = \partial_i\partial_j f(x)$ exists at the origin and is positive definite,
\end{enumerate}
then its asymptotic expansion in the limit $k\to \infty$ can be expressed by
\begin{equation}
  P(k) \sim g(0)\E^{-|k|^sf(0)}\sqrt{\frac{(2\pi)^3}{|k|^{3s}\det A}}
  \exp\left(
    -\frac{k^\top A^{-1}k}{2|k|^s}
  \right)\quad \mathrm{as}\;|k|\to\infty\;.
\end{equation}

We rewrite the non-linear density power spectrum (\ref{final_full_ps}) as
\begin{align}
  \mathcal{P}_\mathrm{free}(k_1,t) &= \int_r\E^{-g_{qp}^2(t)k_1^2\left(
    (m_1(r)-m_1(0))+\mu^2 (m_2(r)-m_2(0))
  \right)}\E^{-\I k_1r\mu} \nonumber\\ &-
  \int_r\E^{g_{qp}^2(t)k_1^2(m_1(0)+\mu^2 m_2(0))}\E^{-\I k_1r\mu}\;.
\end{align}
As will be shown in (\ref{eq:delta_lk_0}), the second term drops to zero exponentially for $k_1\to\infty$, allowing us to restrict the asymptotic analysis to the first of these integrals. The functions $f(x)$ and $g(x)$ from (\ref{eq:81}) become
\begin{equation}
  f(r) = g_{qp}^2(t)\left(
    (m_1(r)-m_1(0))+\left(m_2(r)-m_2(0)\right)\mu^2
  \right)\;,\quad g(r) = 1\;.
\end{equation}
Requirements \ref{C1} and \ref{C2} are obviously satisfied based on the analysis of $m_1(r)$ and $m_2(r)$ in (\ref{a1}) and (\ref{a2}). To examine requirement \ref{C3}, we expand $f(r)$ into a Taylor series around $r = 0$,
\begin{equation}
  f(r) = \left(b_1-b_0-3\mu^2b_1\right)
  \frac{g_{qp}^2(t)}{2^{5/6}\pi^2}\bar{\sigma}^{7/3}
  \Gamma\left(\frac{3}{2}\right)\,
  \sum_{n=1}^\infty
  \frac{\Gamma\left(\frac{7}{6}+n\right)}{\Gamma\left(\frac{3}{2}+n\right)}
  \frac{\left(-\frac{\bar{\sigma}^2r^2}{2}\right)^n}{n!}\;.
\end{equation}

Thus the Hessian $A$ at the origin reduces to the following matrix
\begin{equation}\label{A}
  A = g_{qp}^2\sigma_1^2\left[
    (b_0-b_1)\mathbb{1}_3+3b_1\hat k_1\otimes\hat k_1
  \right]\;,\quad
  \sigma_1^2 = \frac{\bar\sigma^{13/3}}{2^{5/6}\pi^2}\,
  \frac{\Gamma\left(\frac{3}{2}\right)\Gamma\left(\frac{13}{6}\right)}
       {\Gamma\left(\frac{5}{2}\right)}\;,
\end{equation}
with $b_0>b_1>0$. This shows that requirement \ref{C3} is fulfilled as well.

The non-linear density power spectrum $\mathcal{P}_\mathrm{free}(k_1,t)$ in (\ref{final_full_ps}) thus meets all three requirements. Since $f(0) = 0$ and $g(0) = 1$, its asymptotic expansion for $k_1\to \infty$ reads
\begin{equation}\label{p_asymp}
  \mathcal{P}(k_1) \sim \frac{1}{|k_1|^3}\,
  \sqrt{\frac{(2\pi)^3}{\det A}}\exp\left(
    -\frac{k_1^\top A^{-1}k_1}{2|k_1|^2}
  \right) = \frac{\mathcal{P}^{(0)}(t)}{|k_1|^3}\;.
\end{equation}
This $k^{-3}$ asymptotic behaviour only depends on the number of spatial dimensions ($d=3$) and the shape of the function $f(r)$. Since $f(r)$ solely depends on the initial momentum correlation functions of dust particles, the small-scale tail proportional to $k^{-3}$ is entirely fixed by the initial conditions of the particle ensemble. As time passes, the full non-linear density power spectrum will thus always approach the $k^{-3}$ asymptotic behaviour. Let us now explore its time dependence.

\subsubsection{Time dependence}

The right panel of Fig.~\ref{fig:ps_com} shows the non-linear density power spectrum $\mathcal{P}_\mathrm{free}(k_1,t)$ at four different times. All these spectra develop the universal $k^{-3}$ slope at small scales. At larger scales, the spectrum grows with time, while it first increases then decreases at small scales. This reflects that structures form by correlated, freely streaming particles. Their collective streaming builds up structures at all scales first, but then destroys them later on scales small enough for particle streams to cross.

\begin{figure}
  \centering
  \includegraphics[width=0.47\hsize]{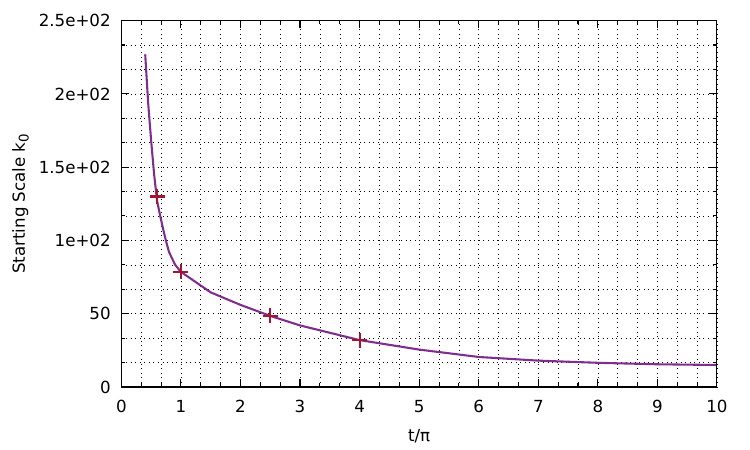}\hfill
  \includegraphics[width=0.47\hsize]{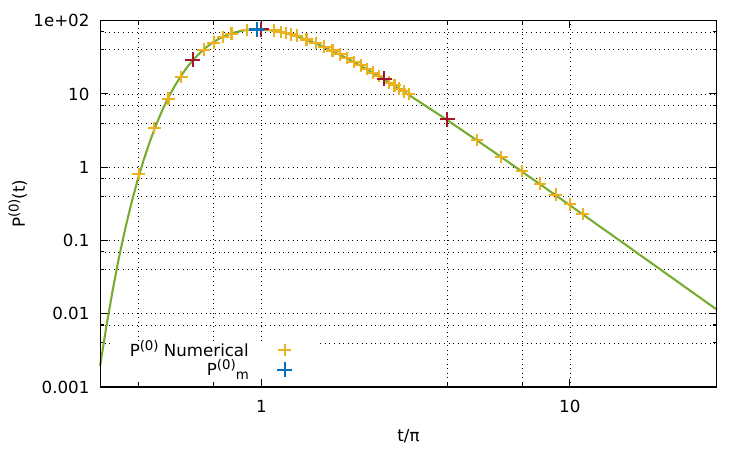}
\caption{\emph{Left}: The wave number $k_0$ where the free spectrum reaches its asymptotic slope decreases with time first rapidly, then slowly, showing how structure formation proceeds towards larger scales. \emph{Right}: Amplitude $\mathcal{P}^{(0)}(t)$ for the asymptotic $k^{-3}$ tail of the free, non-linear power spectrum as a function of time $t$. The blue point indicates its maximum $\mathcal{P}^{(0)}_\mathrm{max}$ at time $t_\mathrm{max}$ according to (\ref{p0ms}). The yellow points are obtained by directly integrating (\ref{final_full_ps}). The four red points mark the amplitudes at the times $t=0.6\pi$, $t=1.0\pi$, $t=2.5\pi$ and $t=4.0\pi$.}
\label{fig:p0t}
\end{figure}

To better understand how the amplitude of the $k^{-3}$ asymptotic tail depends on time, the left panel of Fig.~\ref{fig:p0t} shows how the wave number $k_0$ where the free spectrum reaches its asymptotic slope decreases and thus moves towards larger scales. The decreasing slope of this curve shows that structure formation rapidly proceeds towards small scales first and then slows down. At the same time, the amplitude of the $k^{-3}$ tail first increases, then decreases, as shown in the right panel.

We further evaluate the amplitude $\mathcal{P}^{(0)}$ of the asymptotic tail defined in (\ref{p_asymp}). The Hessian $A$ from (\ref{A}) has the determinant
\begin{equation}
  \det A = \left(b_0-b_1\right)^2\left(b_0+2b_1\right)g_{qp}^6(t)\sigma_1^6
\end{equation}
and the inverse
\begin{equation}
  A^{-1} = \frac{1}{g_{qp}^2(t)\sigma_1^2}
  \frac{1}{b_0-b_1}\left(
    \mathbb{1}_{3}-
    \frac{3b_1}{\left(b_0+2b_1\right)}\hat k_1\otimes\hat{k_1}
  \right)\;,
\end{equation}
thus the amplitude $\mathcal{P}^{(0)}(t)$ becomes
\begin{equation}\label{eq:final_p0t}
  \mathcal{P}^{(0)}(t) = \frac{1}{g_{qp}^3(t)\sigma_1^3\left(b_0-b_1\right)}
  \sqrt{\frac{(2\pi)^3}{b_0+2b_1}}\,\exp\left(
    -\frac{1}{2g_{qp}^2(t)\sigma_1^2\left(b_0+2b_1\right)}
  \right)\;.
\end{equation}
It reaches its maximum
\begin{equation}\label{p0ms}
  \mathcal{P}^{(0)}_\mathrm{max} = \left(\frac{6\pi}{\E}\right)^{3/2}
  \frac{b_0+2b_1}{b_0-b_1} \approx 75.53
  \quad\mbox{at}\quad
  t_\mathrm{max} = \sigma_1^{-1}
  \sqrt{\frac{1}{3\left(b_0+2b_1\right)}} \approx 0.97\pi\;.
\end{equation}

The right panel of Fig.~\ref{fig:p0t} shows the amplitude $\mathcal{P}^{(0)}(t)$ for the $k^{-3}$ asymptotic tail as a function of time, calculated analytically by (\ref{eq:final_p0t}) and numerically by directly integrating (\ref{final_full_ps}). Their perfect match confirms the asymptotic result. The steep increase at early times shows that small-scale structures form rapidly. After $t_\mathrm{max}$, the amplitude of the $k^{-3}$ slope decreases, showing that small-scale structures are destroyed after formerly convergent particle streams cross and begin diverging.

\section{Summary and Discussion} \label{sec:diss}

Kinetic field theory applies the principles and concepts of a statistical field theory to ensembles of classical particles in or out of equilibrium. The particle ensemble is characterized by the probability distribution of its initial phase-space coordinates and by the equations of motion for the particles. An exact and complete generating functional can be defined such that the statistical properties of the particle ensemble at any later time can be derived by applying appropriate functional derivatives to this generating functional. Up to this point, the theory is applicable to wide classes of classical particle ensembles, or, more generally, to ensembles of classical degrees of freedom.

The focus of this paper has been a first application of KFT to planet formation. Compared to numerical simulations of protoplanetary disks, KFT offers the advantage of not being limited by spatial resolution. In contrast to other analytic approaches to planetesimal formation, the equations of motion for dust particles are deterministic and enter the generating functional via their Green's function. We model the initial state for the dust particles at the example of an evolved snapshot of a numerical simulation. Based on these conditions, applying KFT to planetary formation is straightforward.

Our first main result is an initial phase-space distribution of dust particles in phase space which models the auto- and cross-correlations of particle positions and momenta in a way representing a time slice of the simulation. While the momentum components follow a Gaussian distribution, the particle density does not. We have shown however that the square of the particle density is a Gaussian variable at reasonable approximation. The auto- and cross-spectra of the particle momenta and squared densities show a characteristic form, beginning as a power law on large scales and turning into an exponential decrease at small scales. We calculate the corresponding auto- and cross-correlation functions by fast Fourier transforms. Assuming a multivariate Gaussian for particle momenta and squared densities, the initial probability distribution is fixed. We convert it to an initial probability distribution for particle momenta and positions and thus complete the statistical description of the initial state.

Our second main result are the expressions (\ref{z_0}) for the free generating functional $Z_0[\tens L]$ and (\ref{final_full_ps}) for the non-linearly evolved density-fluctuation power spectrum $\mathcal{P}_\mathrm{free}(k_1,t)$. In the free generating functional, particle interactions are neglected. We are thus studying the evolution of freely streaming dust particles, with their phase-space coordinates chosen such that they stream collectively in a correlated manner. The free power spectrum is calculated from $Z_0$ as usual in KFT, i.e.\ by applying two density operators to $Z_0$. The non-linear density-fluctuation power spectrum universally develops a $k^{-3}$ tail at small scales, implying scale-invariant structure formation below a characteristic length scale, which is typically way below the resolution limit of numerical simulations at early times.

Our third main result is the proof that this universal, asymptotic $k^{-3}$ tail develops necessarily. For doing so, we have shown that an asymptotic analysis published elsewhere in the context of cosmic structure formation can be applied to protoplanetary disks as well, and extended it to derive analytic expressions for the amplitude of the asymptotic tail, its evolution with time, and the time scale when it reaches its maximum. The wave number $k_0$ where the asymptotic behaviour sets in decreases with time, first rapidly, then slowly, showing that small-scale structure formation proceeds quickly initially and then slows down. 

When performing numerical simulations it is important to show that the chosen initial random perturbation does not lead to preferred structure formation on certain scales. For our numerical streaming instability experiments we can thus show that based on the KFT analysis the formation of self-similar structures at small scales by free streaming from a generalized unstratified streaming instability initial condition is inevitable under few and very general assumptions, which is an important result for modelers of this process in general. It remains to be seen whether this result will hold up once particle interactions and friction will be taken into account.

So far, we have neglected particle interactions, analyzing the effects of a coherent, free flow of dust particles from initially correlated phase-space positions. Our next step will be to include particle interactions with the surrounding gas and with each other, we expect that in particular mean-field approaches can be developed and applied for particle interactions described by more realistic Hamiltonians.

\section*{Acknowledgement}

It is a pleasure to thank Sara Konrad and Andreas Schreiber for many helpful comments and interesting discussions. This work has been funded by Deutsche Forschungsgemeinschaft (DFG, German Research Foundation) under Germany’s Excellence Strategy EXC-2181/1 - 390900948 (the Heidelberg STRUCTURES Cluster of Excellence).

\section*{Data Availability}

The data that support the findings of this study are available from the corresponding author, J. Shi, upon reasonable request.	

\bibliographystyle{mnras}
\bibliography{references}

\appendix

\section{Concepts of KFT in comparison to equilibrium statistical physics}
\label{app:kft}

In equilibrium thermodynamics, the partition sum $Z$ contains all macroscopic information on a particle ensemble. Taking derivatives of $Z$, expectation values for macroscopic state variables can be retrieved. For example, the (negative) derivative of $Z$ with respect to the inverse temperature $\beta$ returns the mean internal energy $U$, and the derivative of $Z$ with respect to the chemical potential $\mu$ returns the mean particle number $\langle N\rangle$ in a grand-canonical ensemble.

Kinetic field theory (KFT) proceeds in an analogous way: it is based on an equivalent to the partition sum for a classical particle ensemble, from which macroscopic, statistical information is retrieved via derivatives. There is one important extension, though: Since KFT is a theory for systems possibly out of equilibrium, its equivalent to the partition sum needs to depend on time. It therefore needs to be a functional of ensemble properties, which are themselves functions of time. For this reason, $Z$ is called a generating functional in KFT. Likewise, derivatives need to return functions rather than numbers, and therefore need to be functional derivatives. In KFT, time-dependent, macroscopic, statistical information on classical particle ensembles is thus obtained by taking functional derivatives of the generating functional $Z$, conceptually similar to thermodynamics.

In equilibrium thermodynamics, the partition sum is an integral over a probability distribution for microstates. In KFT, the generating functional is an integral over the probability with which phase-space positions at any time $t$ will be reached by particles. This probability is determined by the probability for an initial phase-space position to be occupied by a particle, times the conditional transition probability for this particle from its initial to a later phase-space position.

The probability distribution for the initial phase-space positions characterizes the initial state of the entire ensemble and needs to be suitably chosen, taking possible correlations of particle positions and momenta into account. A common choice is a multivariate Gaussian, motivated by the central limit theorem.

Any initial position in phase space is the starting point of a classical particle trajectory, which solves the equation of motion with the initial phase-space point taken as initial condition. For classical particles, the transition probability from initial to final states takes the form of a (functional) Dirac delta distribution: it vanishes almost everywhere, except along the actual particle trajectory. This is the main difference to quantum mechanics or quantum field theory, where the transition probability is a path integral over a phase factor containing the action functional.

Particle trajectories beginning at arbitrary phase-space points can frequently be characterized by a Green's function which formally solves the equation of motion. This solution has typically two parts, a free part neglecting interactions, and an interacting part containing a time integral over possible forces times the Green's function. The free generating functional $Z_0$ contains only the free part of the trajectories, multiplying the initial phase-space point by the Green's function.

With the Green's function and the initial probability distribution, the generating functional of KFT is completely specified. Statistical properties at any time later than the initial time can then be read off by suitable functional derivatives. Such derivatives are the essential part of operators, for which the density operator is the most common example. Particle interactions can also be introduced by an interaction operator acting on the free generating functional. This interaction operator can be Taylor-expanded, leading to perturbation theory, or averaged, leading to mean-field theory. Therefore, the free generating functional is the heart of KFT. The interaction operator, in any suitable approximation, adds interactions to it. Further operators such as the density operator then extract statistical information. For example, applying two density operators returs a density correlation function.

\section{The Initial Probability Distribution}
\label{appa}

To calculate the full expression of the initial probability distribution of dust particles in phase space, we first introduce the data tensor
\begin{equation}
  \tens d :=
  \begin{pmatrix} \bar{\rho}\\ \Vec{p} \end{pmatrix}_j \otimes \Vec{e}_j
\end{equation}
and its Fourier-conjugate
\begin{equation}
  \tens t :=
  \begin{pmatrix} t_{\bar{\rho}}\\ \Vec{t}_p \end{pmatrix}_j \otimes \Vec{e}_j\;.
\end{equation}
Similarly, the covariance matrix $\Bar{C}$ of the $N$ particles can be decomposed as
\begin{equation}\label{cov_matr}
  \Bar{C} = \left<\tens d\otimes\tens d\right> =
  \begin{pmatrix}
    C_{\bar\rho\bar\rho} & C_{\bar\rho p}^\top \\
    C_{\bar\rho p} & C_{pp}
    \end{pmatrix}=
    \begin{pmatrix}
    \left<\bar{\rho}_j\bar{\rho}_k\right> &
    \left<\bar{\rho}_j\Vec{p}_k\right>^\top \\
    \left<\Vec{p}_j\bar{\rho}_k\right> &
    \left<\Vec{p}_j\otimes\Vec{p}_k\right>
  \end{pmatrix}\otimes E_{jk}
\end{equation}
with $E_{jk}=\Vec{e}_j\otimes\Vec{e}_k$.

Since the positions $\Vec q_j$ of the $N$ particles in configuration space are required to sample the density distribution, we set
\begin{equation}\label{q_z}
  P(\Vec{q}_j|\bar{\rho}_j) =
  \int\D\rho_j P(\Vec{q}_j|\rho_j)P(\rho_j|\bar{\rho}_j) =
  \int\D\rho_j\frac{\rho_j}{N}\,\delta_\mathrm{D}\left(
    \rho_j-\rho_j(\bar{\rho}_j)
  \right) = \frac{\rho_j(\bar{\rho}_j)}{N}\;,
\end{equation}
assuming Poisson sampling. The probability for finding a particle at position $\Vec{q}_j$ with momentum $\Vec{p}_j$ can thus be expressed by
\begin{equation}
  P(\Vec{q}_j,\Vec{p}_j) = \int\D\bar{\rho}_j\,
  P(\Vec{q}_j|\bar{\rho}_j)\,P(\bar{\rho}_j,\Vec{p}_j) =
  N^{-1}\int\D\bar{\rho}_j\rho_j(\bar{\rho}_j)P(\bar{\rho}_j,\Vec{p}_j)\;,
\end{equation}
and the probability distribution for the complete set $\{\Vec{q}_j,\Vec{p}_j\}$ of $N$ phase-space coordinates is
\begin{equation}\label{full_prob_distri}
  P\left(\{\Vec{q}_j,\Vec{p}_j\}\right) = N^{-N}\int\D^{N}\bar{\rho}
  \prod_{j=1}^{N}\rho_j(\bar{\rho}_j)\,P(\{\bar{\rho}_j,\Vec{p}_j\}) =
  N^{-N}\int\D^{N}\bar{\rho}\prod_{j=1}^{N}\bar{\rho}_j^2\,P(\tens d)\;.
\end{equation}

For evaluating $P(\tens d)$, we turn to its characteristic function
\begin{equation}\label{ch_dp}
  \phi(\tens t) = \exp\left(
    -\frac{1}{2}\tens t^\top\Bar{C}\tens t+
    \I\tens t^\top\tens\mu
  \right)\;,
\end{equation}
where $\tens\mu=\left(\mu_{\bar\rho},\vec \mu_p\right)^T_j\otimes \vec e_j$ contains the mean values of the Gaussian variables. The probability distribution for the data tensor is then given by the inverse Fourier transform of the characteristic function
\begin{align}
  P(\tens d) &= \int\frac{\D\tens t_p}{(2\pi)^{3N}}
  \exp\left(
    -\frac{1}{2}\tens t_p^\top\Bar{C}_{pp}\tens t_p+
    \I\tens t_p^\top\tens\mu_p+
    \I\left<\tens t_p,\tens p\right>
  \right) \nonumber\\ &\cdot
  \int\frac{\D\tens t_{\bar{\rho}}}{(2\pi)^N}
  \exp\left(
    -\frac{1}{2}\tens t_{\bar{\rho}}^\top\Bar{C}_{\bar{\rho}\bar{\rho}}
    \tens t_{\bar{\rho}}-
    \tens t_{\bar{\rho}}^\top\Bar{C}_{\bar{\rho} p}\tens t_p+
    \I\tens t_{\bar{\rho}}^\top\tens\mu_{\bar{\rho}}+
    \I\left<\tens t_{\bar{\rho}},\tens{\bar{\rho}}\right>
  \right)\;,
\end{align}
where we have defined the tensors
\begin{equation}
  \tens t_{\bar{\rho}} := t_{\bar{\rho}_j}\otimes \Vec{e}_j\;,\quad
  \tens t_p := t_{p_j}\otimes \Vec{e}_j\;,\quad
  \tens{\bar{\rho}} := \bar{\rho}_j\otimes\Vec{e}_j\;,\quad
  \tens p := \Vec{p}_j\otimes\Vec{e}_j\;,\quad
  \tens{\mu_{\bar{\rho}}} := \mu_{\bar{\rho}_j}\otimes\Vec{e}_j\;,\quad
  \tens \mu_p := \vec \mu_{p_j}\otimes\Vec{e}_j
\end{equation}
and the covariance matrices
\begin{equation}
  \Bar{C}_{\bar{\rho}\bar{\rho}} := \Bar{C}_{\bar{\rho}_j\bar{\rho}_k}\otimes E_{jk}\;,\quad
  \Bar{C}_{\bar{\rho} p}:=\Bar{C}_{\bar{\rho}_jp_k}\otimes E_{jk}\;,\quad
  \Bar{C}_{pp}:=\Bar{C}_{p_jp_k}\otimes E_{jk}\;.
\end{equation}
The full probability distribution for the phase-space coordinates of the $N$ particles is
\begin{align}
  P(\tens q,\tens p) &= P(\{\Vec{q}_j,\Vec{p}_j\}) \nonumber\\ &=
  N^{-N}\int\D\tens{\bar{\rho}} \prod_{j=1}^N\bar{\rho}_j^2
  \int\frac{\D\tens t_p}{(2\pi)^{3N}}\exp\left(
    -\frac{1}{2}\tens t_p^\top\Bar{C}_{pp}\tens t_p+
    \I\tens t_p^\top\tens\mu_p+
    \I\left<\tens t_p,\tens p\right>
  \right) \nonumber\\ &\cdot
  \int\frac{\D\tens t_{\bar{\rho}}}{(2\pi)^N}\exp\left(
    -\frac{1}{2}\tens t_{\bar{\rho}}^\top\Bar{C}_{\bar{\rho}\bar{\rho}}
    \tens t_{\bar{\rho}}-
    \tens t_{\bar{\rho}}^\top\Bar{C}_{\bar{\rho} p}\tens t_p+
    \I\tens t_{\bar{\rho}}^\top\tens\mu_{\bar{\rho}}+
    \I\left<\tens t_{\bar{\rho}},\tens{\bar{\rho}}\right>
  \right)\;.
\end{align}
We first integrate over $\tens{\bar{\rho}}$ using
\begin{equation}
  \int\D\bar\rho_j\,\bar\rho_j^2\,\exp\left(
    \I t_{\bar\rho_j}\,\bar\rho_j
  \right) = -\frac{\partial^2}{\partial t_{\bar\rho_j}^2}\int\D\bar\rho_j\,
  \exp\left(\I t_{\bar\rho_j}\,\bar\rho_j\right) =
  -2\pi\frac{\partial^2}{\partial t_{\bar\rho_j}^2}
  \delta_\mathrm{D}\left(t_{\bar\rho_j}\right)\;.
\label{eq:delta}
\end{equation}
Integrating twice by parts with respect to $t_{\bar\rho_j}$, and using the delta distribution just obtained, we arrive at
\begin{equation}\label{I_3_int}
  P(\tens q,\tens p) = N^{-N}\int\frac{\D\tens t_p}{(2\pi)^{3N}}\,
  \mathcal{C}(\tens t_p)\exp\left(
    -\frac{1}{2}\tens t_p^\top\Bar{C}_{pp}\tens t_p+
    \I\left<\tens t_p,\tens p\right>
  \right)
\end{equation}
with
\begin{equation}
  \mathcal{C}(\tens t_p) = (-1)^N\left(
    \prod_{j=1}^{N}\frac{\partial^2}{\partial t_{\bar{\rho}_j}^2}
  \right)\left.\exp\left(
    -\frac{1}{2}\tens t_{\bar{\rho}}^\top\Bar{C}_{\bar{\rho}\bar{\rho}}\tens t_{\bar{\rho}}-
    \tens t_{\bar{\rho}}^\top\Bar{C}_{\bar{\rho} p}\tens t_p+
    \I\tens t_{\bar{\rho}}^\top\tens\mu_{\bar{\rho}}
  \right)\right|_{\tens t_{\bar{\rho}}=0}\;.
\end{equation}
And we've used $\tens \mu_p=0$ for ignoring the bulk velocity of particles. Notice that the delta distribution in (\ref{eq:delta}) requires the integration range to be infinite. While this is not strictly the case for the density variable $\bar{\rho}$, the integration range is still wide enough for the integral to approximate the delta distribution well.

The expression $\mathcal{C}(\tens t_p)$ consists of terms expressing a hierarchy of products of correlation functions,
\begin{align}\label{eq:ctp}
  \mathcal{C}(\tens t_p) &= \prod_{j=1}^{N}\left(
    \Bar{C}_{\bar{\rho}_j\bar{\rho}_j}-\left(
      -\Bar{C}_{\bar{\rho}_j p}\tens t_{p}+\I\mu_{\bar{\rho}_j}
    \right)^2
  \right) \nonumber\\ &+
  \sum_{(j,k)}2\Bar{C}_{\bar{\rho}_j\bar{\rho}_k}\left(
    \Bar{C}_{\bar{\rho}_j\bar{\rho}_k}-2\left(
      -\Bar{C}_{\bar{\rho}_jp}\tens t_{p}+\I\mu_{\bar{\rho}_j}
    \right)\left(
      -\Bar{C}_{\bar{\rho}_kp}\tens t_{p}+\I\mu_{\bar{\rho}_k}
    \right)
  \right)\prod_{\{l\}}\left(
    \Bar{C}_{\bar{\rho}_l\bar{\rho}_l}-\left(
      -\Bar{C}_{\bar{\rho}_l p}\tens t_{p}+\I\mu_{\bar{\rho}_l}
    \right)^2
  \right) \nonumber\\ &+ \ldots\;.
\end{align}
In the second line, the sum extends over all pairs $(j,k \ne j)$, and the product includes all indices $l$ except $(j, k)$. We convert $\mathcal{C}(\tens t_p)$ into an operator by replacing $\tens t_p\to-\I\partial_{\tens p}$, pull it out of the integral in (\ref{I_3_int}) and carry out the final Gaussian integration to arrive at
\begin{equation}\label{final_prob_dist_app}
  P(\tens q,\tens p) = \frac{N^{-N}\mathcal{C}(-\I\partial_{\tens p})}
  {\sqrt{(2\pi)^{3N}\det\Bar{C}_{pp}}}\exp\left(
    -\frac{1}{2}\tens p^\top\Bar{C}_{pp}^{-1}\tens p
  \right)\;.
\end{equation}

\section{Covariance Matrix}\label{pip_CM}

\subsection{Correlation functions and Power spectrums}\label{cf_ps}

Corresponding to the three types of covariance matrices $\bar{C}_{pp}$, $\bar{C}_{\bar{\rho}\bar{\rho}}$ and $\bar{C}_{\bar{\rho} p}$, there are three sets of spatial correlation functions, which respectively are density, momentum, and density-momentum correlation functions. Denoting density as $\bar{\rho}$ and momentum as $p_i$ for $i=\{1,2,3\}$, the density correlation function is defined as
\begin{equation}\label{df_den_cf}
  \zeta_{00}(\Vec{r}) = \left<\bar{\rho}(\Vec{q})\bar{\rho}(\Vec{q}+\Vec{r})\right> =
  \frac{1}{L^3}\int_{q}\bar{\rho}(\Vec{q})\bar{\rho}(\Vec{q}+\Vec{r})\;.
\end{equation}
In the $N$-body simulation, $L$ is the size of the simulation box. The spatial correlation functions of the momenta follow the identical concept, but the momenta are three-dimensional. We define
\begin{equation}\label{df_mom_cf}
  \zeta_{ij}(\Vec{r}) = \left<p_i(\Vec{q})p_j(\Vec{q}+\Vec{r})\right> =
  \frac{1}{L^3}\int_{q} p_i(\Vec{q})p_j(\Vec{q}+\Vec{r}) \;,\quad
  i,j=1,2,3\;.
\end{equation}
They describe how each of the momentum components statistical relates to each other at different positions. Since covariance matrices are all symmetrical, there are $n_{pp}=6$ momentum correlation functions in three dimensions.

The spatial cross-correlation functions between density and momentum components are defined similarly by
\begin{equation}\label{df_dm_cf}
  \zeta_{0i}(\Vec{r}) = \left<\bar{\rho}(\Vec{q})p_i(\Vec{q}+\Vec{r})\right> =
  \frac{1}{L^3}\int_{q} \bar{\rho}(\Vec{q})p_i(\Vec{q}+\Vec{r})\;,\quad
  i=1,2,3\;.
\end{equation}
They describe how the density relates to each of the momentum components at different positions. In three dimensions, there are $n_{\bar{\rho} p}=3$ spatial density-momentum cross-correlation functions.

Power spectra are the Fourier transforms of the spatial correlation functions defined as $P(\vec{r})=\int_r \zeta(\vec{r})e^{-i\vec{k}\cdot\vec{r}}$. Denoting the Fourier transform of the density by $\bar{\rho}_k$ and of the momentum as $p_{ki}$ for $i=\{1,2,3\}$, we can compute the spatial power spectra for (\ref{df_den_cf}), (\ref{df_mom_cf}) and (\ref{df_dm_cf}) as
\begin{equation}\label{df_ps}
  P_{00}(\Vec{k}) = \frac{1}{L^3} \bar{\rho}_k(\Vec{k})\bar{\rho}_k(-\Vec{k})\;,\quad
  P_{ij}(\Vec{k}) = \frac{1}{L^3} p_{kj}(\Vec{k})p_{ki}(-\Vec{k})\;,\quad
  P_{0i}(\Vec{k}) = \frac{1}{L^3} p_{ki}(\Vec{k})\bar{\rho}_k(-\Vec{k})\;.
\end{equation}

\subsection{The pipeline}\label{pipline}

Due to the enormous number of dust particles, we have chosen a different path than computing their spatial correlation functions directly. The essential idea is that by transforming the density and momenta fields into Fourier space, we will initially calculate their spatial power spectra defined in (\ref{df_ps}), then fit appropriate functions to the numerical power spectra, and obtain our spatial correlation functions and thus the components of the covariance matrices by performing simple analytical integrations.

We perform the Fourier transform by FFTW. This requires us to sort the particles into grid cells first. We do so by applying the cloud-in-cell (CIC) method, which assigns the mass of each particle in the simulation box to all grid cells next to it (in the three-dimensional case, the number of neighbouring cells is $2^3=8$).

For transforming momenta to Fourier space, we approximate the momentum of each cell by the momentum of the particle closest to its centre. This is justified if particle momenta do not change appreciably across a cell. We set the number of grid cells to 256 per dimension, hence each cell contains on average $\frac{N}{256^3}=1.25\approx 1$ particles. For this reason, we consider our method of extracting momenta information to be appropriate.

From the Fourier transforms of density and momentum fields, we could directly calculate their power spectra based on (\ref{df_ps}). To further simplify the algorithm, if we assume that the simulation box is substantially larger than all correlation scales, the simulated particle distribution can be considered as isotropic. Then, the power spectra we are looking for will not depend on the direction of the wave vector $\vec{k}$ but only on the wave number. We can then write the density power spectrum as
\begin{equation}\label{ps_den_final}
  \bar P_{00}(k) = P_{00}(\vec k) =
  \frac{1}{L^3}\bar{\rho}_k(\Vec{k})\bar{\rho}_k(-\Vec{k}) =
  \frac{1}{L^3}|\bar{\rho}_k(\Vec{k})|^2\;.
\end{equation}

Since the covariance of the momenta is symmetric, we define for the momentum power spectra $P_{ij}(\vec{k})$
\begin{equation}\label{ps_mom_final}
  \bar{P}_{ij}(k) = \frac{1}{2}\left(P_{ij}(\vec{k})+P_{ji}(\vec{k})\right) =
  \frac{1}{L^3}\left|
    \Re\left(p_{kj}(\vec{k})\right)\Re\left(p_{ki}(\vec{k})\right)+
    \Im\left(p_{kj}(\vec{k})\right)\Im\left(p_{ki}(\vec{k})\right)
  \right|\;,
\end{equation}
where $\Re(z)$ and $\Im(z)$ are the real and imaginary parts of the complex number $z$. In analogy to the above calculation, we immediately find the new density-momentum cross power spectra
\begin{align}\label{ps_dm_final}
  \bar{P}_{0i}(k) &= \frac{1}{2}\left(P_{0i}(\vec{k})+P_{i0}(\vec{k})\right) =
  \frac{1}{L^3}\left|
    \Re\left(p_{ki}(\vec{k})\right)\Re\left(\bar{\rho}_k(\vec{k})\right)+
    \Im\left(p_{ki}(\vec{k})\right)\Im\left(\bar{\rho}_k(\vec{k})\right)
  \right|\;.
\end{align}
Thus the combined spatial correlation functions $\zeta_{\mu\nu}(r)$ of density, momenta, and density-momentum are obtained by Fourier transform via
\begin{align}\label{comb_cf}
  \zeta_{\mu\nu}(r) &=\int_k\bar P_{\mu\nu}(k)
  \E^{2\pi\I\vec{k}\cdot\vec{r}} =
  \int_0^\infty\frac{\D k}{2\pi^2}k^2\bar P_{\mu\nu}(k)j_0(2\pi kr)
\end{align}
where $j_0(2\pi kr) = \frac{\sin(2\pi kr)}{2\pi kr}$ is the zeroth-order spherical Bessel function of the first kind; $\mu,\nu=0,1,2,3$.

\section{Further discussion of the free KFT power spectrum}
\label{app3:more_ps}

Now we return to the full expression in (\ref{eq:new_non_p}), where we denote the second integral as $\mathcal{P}_\mathrm{diff}(k_1,t)$,
\begin{equation}
  \mathcal{P}_\mathrm{diff}(k_1,t) =
  \int_r\E^{g_{qp}^2(t)k_1^2(m_1(0)+\mu^2 m_2(0))}\E^{-\I k_1r\mu}\;.
\end{equation}
First, we analyse the asymptotic behaviour of $\mathcal{P}_\mathrm{diff}(k_1,t)$ at large and small $k_1$. Using the definition of $y(t, \mu)$ in (\ref{y(t,mu)}), when $k_1\to \infty$, at finite evolution time $t$, $\mathcal{P}_\mathrm{diff}(k_1,t)$ becomes
\begin{align}\label{eq:delta_lk_0}
  \lim_{k_1\to\infty}\mathcal{P}_\mathrm{diff}(k_1,t) &=
  \lim_{k_1\to\infty}\int_r\E^{y(t,\mu)k_1^2}\E^{-\I k_1r\mu} =
  \lim_{k_1,R\to\infty}2\pi\int_{-1}^{1}\D\mu
  \E^{y(t,\mu)k_1^2}\int_{0}^{R}\D r\,r^2\cos(k_1 r \mu) \\ &<
  \lim_{k_1,R\to \infty}2\pi R^3\int_{-1}^{1}\D\mu
  \E^{y(t,\mu)k_1^2} < \lim_{k_1,R\to \infty}4\pi R^3\int_{0}^{1}\D\mu
  \E^{y(t,0)k_1^2} \\ &=
  \lim_{k_1,R\to \infty}4\pi R^3e^{y(t,0)k_1^2}\to 0\;,
\end{align}
where $R$ is the upper integral bound. Since the exponential function increases faster than polynomials, $\mathcal{P}_\mathrm{diff}(k_1,t)$ will tend to zero for large $k_1$. On the other hand, if $k_1\to 0$, it becomes
\begin{align}\label{eq:delta_sk_0}
  \lim_{k_1\to 0}\mathcal{P}_\mathrm{diff}(k_1,t) &=
  \lim_{k_1\to 0}\int_r\E^{y(t,\mu)k_1^2}\E^{-\I k_1r\mu} \\ &=
  \lim_{k_1\to 0,R\to\infty}2\pi\int_{0}^{R}\D rr^2
  \int_{-1}^{1}\D\mu\,\cos(k_1 r \mu) \\ &=
  \lim_{k_1\to 0,R\to \infty}4\pi\int_{0}^{R} dr r^2j_0(k_1 r)\to\infty\;,
\end{align}
where $j_0(k_1 r)$ is the zeroth-order spherical Bessel function of the first kind.

Now that the asymptotic behaviour of $\mathcal{P}_\mathrm{diff}(k_1,t)$ is clear, we perform the analytical integral
\begin{equation}
  \mathcal{P}_\mathrm{diff}(k_1,t)=2\pi\int_{0}^{\infty}dr r^2l_0(r)
\end{equation}
to further study how this behaviour changes with different scales. With
\begin{align}
  l_0(r) &= \frac{\sqrt{\pi}\exp\left(
    g_{qp}^2(t)k_1^2m_1(0)+
    \frac{k_1^2r^2}{4g_{qp}^2(t)k_1^2m_2(0)}
  \right)}{2\sqrt{g_{qp}^2(t)k_1^2m_2(0)}} \\ &\cdot \left[
    \mathrm{Erf}\left(
      \frac{k_1r-2\I g_{qp}^2(t)k_1^2m_2(0)}{2\sqrt{g_{qp}^2(t)k_1^2m_2(0)}}
    \right)-\mathrm{Erf}\left(
      \frac{k_1r+2\I g_{qp}^2(t)k_1^2m_2(0)}{2\sqrt{g_{qp}^2(t)k_1^2m_2(0)}}
    \right)
  \right]\;.
\end{align}

\begin{figure}
  \centering
  \includegraphics[width=0.6\hsize]{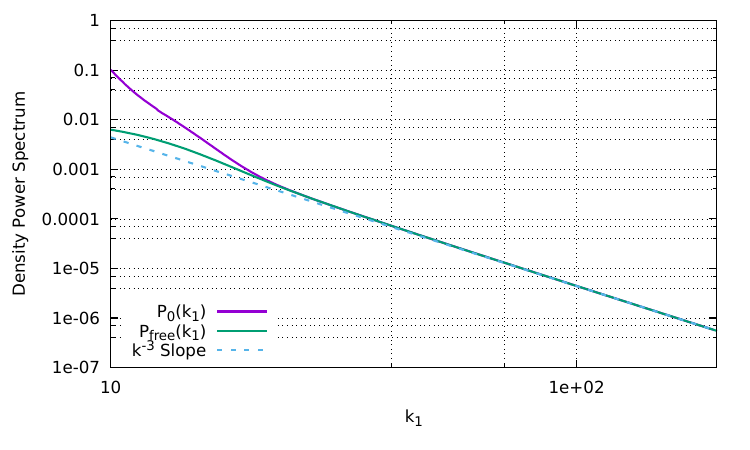}
\caption{This plot compares the non-linear density power spectrum $\mathcal{P}_{0}(k_1,t)$ (purple line) with $\mathcal{P}_\mathrm{free}(k_1,t)$ (green line) together with their asymptotic behaviour at large scales (blue dashed line) at evolution time $t=4\pi$.}
\label{fig:ps2}
\end{figure}

Now to illustrate the effect of neglecting $\mathcal{P}_\mathrm{diff}(k_1,t)$ on the non-linear KFT density power spectrum, we return to its full expression in (\ref{ori_full_ps}) and directly calculate the integral at relative large $k_1$ which guarantees finite results. Figure \ref{fig:ps2} shows the non-linear KFT density power spectrum $\mathcal{P}_{0}(k_1,t)$ and $\mathcal{P}_\mathrm{free}(k_1,t)$ with their small-scale $k^{-3}$ asymptotic behaviour at evolution time $t=4\pi$. At smaller $k_1$, the value of $\mathcal{P}_{0}(k_1,t)$ is indeed larger than $\mathcal{P}_\mathrm{free}(k_1,t)$ and shows rapidly increasing sign. However, at large $k_1$, there is barely any difference between the two curves and they merge very well, which indicates that $\mathcal{P}_\mathrm{diff}(k_1,t)$ does not contribute to nor affect the KFT density power spectrum at small scales and the $k^{-3}$ result remains.

The relative difference between the two curves becomes extremely small above $k\approx45$. Since we are much more interested in the small-scale behaviour, we can safely ignore $\mathcal{P}_\mathrm{diff}(k_1,t)$ in the non-linear KFT density power spectrum without compromising our main results.

% Don't change these lines
\bsp	% typesetting comment
\label{lastpage}
\end{document}